\documentclass[11pt,a4paper]{article}
\usepackage[T1]{fontenc}
\usepackage{lmodern,microtype}
\usepackage[margin=2.5cm]{geometry}
\usepackage{amsmath,amssymb,amsthm,mathtools}
\usepackage{booktabs,tabularx,array,enumitem,xcolor,graphicx}
\usepackage[section]{placeins}
\usepackage{caption}
\usepackage{hyperref}
\hypersetup{colorlinks=true,linkcolor=blue!40!black,citecolor=blue!40!black,urlcolor=blue!40!black,pdftitle={Smoothed Conformal Testing with Small Calibration Sets}}
\usepackage[]{natbib}
\newtheorem{Thm}{Theorem}
\newtheorem{Coro}{Corollary}
\newtheorem{Pro}{Proposition}
\theoremstyle{definition}
\theoremstyle{remark}\newtheorem{remark}{Remark}
\newcommand{\PP}{\mathbb{P}}
\newcommand{\EE}{\mathbb{E}}
\newcommand{\RR}{\mathbb{R}}
\newcommand{\FDR}{\operatorname{FDR}}
\newcommand{\FDP}{\operatorname{FDP}}

\newcommand{\Var}{\operatorname{Var}}
\newcommand{\Unif}{\operatorname{Unif}}
\newcommand{\ind}{\mathbf{1}}
\newcommand{\bC}{\mathbf{C}}
\newcommand{\bP}{\mathbf{P}}
\newcommand{\bQ}{\mathbf{Q}}
\newcommand{\bS}{\mathbf{S}}
\newcommand{\bU}{\mathbf{U}}
\newcommand{\bZ}{\mathbf{Z}}
\newcommand{\cR}{\mathcal{R}}
\newcommand{\cH}{\mathcal{H}}
\newcommand{\cD}{\mathcal{D}}
\newcommand{\cG}{\mathcal{G}}
\newcommand{\cN}{\mathcal{N}}
\newcommand{\cE}{\mathcal{E}}
\newcommand{\cT}{\mathcal{T}}
\newcommand{\Qord}{\bQ^{\mathrm{ord}}}
\setlist{itemsep=2pt,topsep=4pt}
\allowdisplaybreaks
\title{\bfseries Smoothed Conformal Testing with Small Calibration Sets}
\author{%
Mingzhou Deng\thanks{Email: \texttt{dengmingzhou@amss.ac.cn}}\quad
Yan Fu\thanks{Email: \texttt{yfu@amss.ac.cn}}\\[6pt]
\normalsize State Key Laboratory of Mathematical Sciences,\\
\normalsize Academy of Mathematics and Systems Science, Chinese Academy of Sciences%
}
\date{\today}
\begin{document}
\maketitle
\begin{abstract}
A small calibration sample restricts split conformal $p$-values to a coarse rank grid. Randomization removes this discretization, but its effect depends on both the smoothing interval and the allocation of auxiliary random variables. Independent tie-block randomization is exactly valid under null exchangeability. When calibration and null scores are jointly exchangeable conditional on the alternative scores, we prove positive regression dependence and hence Benjamini--Hochberg false discovery rate control, including for atomic score distributions. Smoothing over calibration-dependent rank cells agrees with tie-block randomization when calibration scores are distinct and no test score ties a calibration score, but can fail marginal validity with atoms. Assigning sorted random draws according to score order can invalidate even a single null $p$-value. An exact two-test example violates the usual $\pi_0\alpha$ bound, and simulations with continuous scores exhibit false discovery rates above the nominal level $\alpha$. Finally, conditional moment identities and normal-model illustrations show that removing the rank grid does not remove calibration uncertainty. Randomization can produce additional discoveries, including when ordinary conformal testing cannot reject at all, but it does not generally substitute for a larger calibration sample.
\end{abstract}
\vspace{0.4em}

\section{Introduction}\label{sec:intro}
Multiple hypothesis testing is a fundamental problem in statistics. It concerns controlling an appropriate measure of Type I error across many simultaneous tests. For $m$ hypotheses, let $\cH_0$ index the true nulls, $R$ count all rejections, and $V$ count rejected true nulls. The false discovery proportion and false discovery rate (FDR) are defined by \citet{bh1995}
\[
 \FDP=\frac{V}{R\vee1},\qquad
 \FDR=\EE\!\left[\frac{V}{R\vee1}\right].
\]
Write $p_{(1)}\le\cdots\le p_{(m)}$ for the order statistics of the $p$-value vector.
The Benjamini--Hochberg (BH) procedure is a step-up procedure which rejects
\[
\cR(\mathbf{p}) = \{j : p_j \le \alpha \tau/m\},\quad\text{where }\tau=\max\left(\{k\in\{1,\ldots,m\}:p_{(k)}\le\alpha k/m\}\cup\{0\}\right).
\]
It controls FDR at $\pi_0\alpha$, where $\pi_0=|\cH_0|/m$, under independence \citep{bh1995} or positive regression dependence on the true nulls (PRDS) \citep{by2001}. Broader dependence conditions and operating characteristics are studied by \citet{blanchard2008} and \citet{genovese2002}.

Split conformal inference trains the score function on training data and calibrates it on a held-out calibration set \citep{shafer2008,vovk2012,lei2018,angelopoulos2023}. With calibration scores $C_1,\ldots,C_n$ representing the null distribution and larger test scores $S_j$ indicating stronger evidence, the ordinary conformal $p$-value is
\[
 P_j=\frac{1+\sum_{i=1}^n\ind\{C_i\ge S_j\}}{n+1}.
\]
Conditional on training, exchangeability of the calibration scores and a null test score gives $\PP(P_j\le t)\le t$, ensuring single-test Type I error control \citep{vovk2022,bates2023}. When the pooled calibration scores and null test score are almost surely distinct, the null conformal $p$-value is uniform on the rank grid $\{1/(n+1),\ldots,1\}$. With ties, it is generally conservative. This validity guarantee averages over calibration and test data and need not hold given the realized calibration sample \citep{vovk2012,bian2023}.

For small $n$, the floor $1/(n+1)$ can prevent rejection at multiple-testing thresholds. Related resolution issues occur in Monte Carlo permutation tests \citep{phipson2010}. Uniform randomization fills rank gaps, but its consequences for dependence and power require separate analysis. In particular, independent auxiliary draws do not make the resulting values independent, since all tests still use the same calibration sample. Marginal uniformity alone therefore cannot justify applying BH. Shared-calibration dependence and PRDS are addressed by \citet{bates2023}. Classical tie-block smoothing produces exact null uniformity for a single test under exchangeability, including with atoms \citep{vovk2022}. This single-test guarantee does not by itself establish FDR control for multiple testing. Randomized values for discrete and nonparametric multiple testing are studied by \citet{habiger2011}.

We distinguish two choices, the randomization interval and the allocation of auxiliary uniforms. Tie-block randomization is exactly valid and satisfies PRDS under joint exchangeability given alternative scores, so BH controls FDR with atoms. Calibration-cell randomization sets its interval from the ordinary value and the calibration sample. This removes interval ambiguity when only ranks are retained, but marginal validity can fail with atoms. The two rules coincide whenever calibration scores are distinct and no test score equals a calibration score; these conditions hold almost surely under independent continuous sampling. Under the same no-tie conditions, sorted allocation preserves the BH rejection count, but it can change which hypotheses are rejected. A two-test example violates null validity and the $\pi_0\alpha$ bound, and a larger experiment exceeds $\alpha$. A Simes calculation attributes this to cell occupancy and shows that rejection frequency and identity differ. We also derive probabilities of ignition (an increase in the BH rejection count) by factoring crossing events over rank cells. Identities for the empirical cumulative distribution function (ECDF) and conditional moments show that smoothing retains cell occupancies and calibration uncertainty. Normal-model experiments show higher single-test rejection frequencies, but not full oracle performance or general BH power gains.

Section~\ref{sec:setup} defines tie-block randomization, proves exact marginal validity, and establishes PRDS under joint exchangeability, giving BH FDR control with atoms. It also contrasts calibration-cell randomization with an atomic counterexample. Section~\ref{sec:failure} shows that ordered allocation preserves the BH rejection count under the stated no-tie conditions but can break validity, with Simes identities explaining the mechanism. Section~\ref{sec:ignition} derives conditional ignition probabilities. Section~\ref{sec:limits} separates removal of the rank floor from recovery of null-tail information. Section~\ref{sec:numerics} reports normal-model experiments, and the appendices contain proofs and additional experiments.

\section{Tie-block randomization and FDR control}\label{sec:setup}\label{sec:fdr}
\subsection{Definition and exact validity}
Let $\bC=(C_1,\ldots,C_n)$ be the calibration score vector and $\bS=(S_1,\ldots,S_m)$ the test score vector. We use bold symbols for vectors and ordinary symbols for their scalar components. A larger score indicates stronger evidence against a null hypothesis. The score function is fixed throughout the analysis. If it is fitted using separate training data, the assumptions below are understood conditionally on that training data.

Write $\cH_0\subseteq\{1,\ldots,m\}$ for the true-null set and $\pi_0=|\cH_0|/m$. Define
\begin{equation}\label{eq:ranks}
 A_j=\sum_{i=1}^n\ind\{C_i>S_j\},\qquad E_j=\sum_{i=1}^n\ind\{C_i=S_j\},\qquad
 P_j=(A_j+E_j+1)\delta,\quad \delta=\frac1{n+1}.
\end{equation}
Write $\bP=(P_1,\ldots,P_m)$. For each $j\in \cH_0$, we assume that $(C_1,\ldots,C_n,S_j)$ is exchangeable whenever marginal conformal validity is invoked. Alternative scores need not satisfy this condition. If the pooled scores are distinct almost surely, $P_j$ is uniform on the rank grid $\{\delta,2\delta,\ldots,1\}$ under the null. With ties, it is generally conservative.

Let $\bU=(U_1,\ldots,U_m)$ consist of independent $\Unif(0,1)$ variables, independent of all scores. We write $\PP_{\bU}$, $\EE_{\bU}$, and $\Var_{\bU}$ for probability, expectation, and variance over $\bU$, conditional on all scores. Classical tie-block smoothing \citep{shafer2008,vovk2022} is
\begin{equation}\label{eq:tb}
 Q_j^{\mathrm{tb}}=\delta\{A_j+(E_j+1)U_j\}.
\end{equation}
We collect these values in $\bQ^{\mathrm{tb}}=(Q_1^{\mathrm{tb}},\ldots,Q_m^{\mathrm{tb}})$. Conditional on the scores, each $Q_j^{\mathrm{tb}}$ is uniform on $(A_j\delta,P_j)$. The interval accounts for all calibration scores tied with $S_j$, together with the test score itself.

\begin{Pro}[Marginal exactness]\label{prop:tb}
If $(C_1,\ldots,C_n,S_j)$ is exchangeable and $U_j$ is independent and uniform, then $Q_j^{\mathrm{tb}}\sim\Unif(0,1)$.
\end{Pro}

Proposition~\ref{prop:tb}, proved in Appendix~\ref{app:tb}, includes atomic score distributions. It concerns one null coordinate. To establish the dependence property needed for BH, we next impose a joint exchangeability condition.

\subsection{PRDS and BH FDR control}
For a random vector $\bQ=(Q_1,\ldots,Q_m)$, positive regression dependence on the true nulls, abbreviated PRDS \citep{by2001}, means that for every coordinatewise increasing measurable set $D\subseteq[0,1]^m$ and every $i\in \cH_0$, the conditional probability
\[
 x\longmapsto\PP(\bQ\in D\mid Q_i=x)
\]
admits a nondecreasing version on the support of $Q_i$. An increasing set contains $\mathbf{y}$ whenever it contains $\mathbf{x}$ and $\mathbf{x}\le \mathbf{y}$ coordinatewise. Conditional probabilities at individual values of a continuous coordinate are understood through a regular conditional distribution.

Marginal exchangeability is sufficient for Proposition~\ref{prop:tb}. For the joint result, we impose the stronger condition
\begin{equation}\label{eq:joint-exchangeability}
 \bigl(\bC,\bS_{\cH_0}\bigr)
 \;\big|\; \bS_{\cH_0^c}
 \quad\text{is jointly exchangeable almost surely}.
\end{equation}
Here $\bS_{\cH_0}$ and $\bS_{\cH_0^c}$ are the subvectors of null and alternative scores. The condition requires invariance under every permutation of the calibration and null coordinates, conditional on all alternative scores. It allows discrete, continuous, and mixed score distributions. One sufficient condition is that the calibration and null scores are i.i.d.\ from a common distribution, with the alternative score vector independent of them. Under the complete null, (\ref{eq:joint-exchangeability}) reduces to joint exchangeability of all calibration and test scores.

\begin{Thm}[PRDS of independent tie-block randomization]\label{thm:transfer}
Assume (\ref{eq:joint-exchangeability}), and let the uniforms in (\ref{eq:tb}) be mutually independent and independent of all scores. Then $\bQ^{\mathrm{tb}}$ is PRDS on $\cH_0$. The same conclusion holds conditional on the alternative score vector, for almost every realization of that vector. No continuity assumption is required.
\end{Thm}

Appendix~\ref{app:transfer} proves the theorem by a coupling that remains monotone with ties. Independent auxiliary draws leave dependence through the shared calibration sample; the randomized coordinates need not be independent.

For any vector $\mathbf{p}=(p_1,\ldots,p_m)\in[0,1]^m$, let
\begin{equation}\label{eq:bh}
 R(\mathbf{p})=\max\left(\{k\in\{1,\ldots,m\}:p_{(k)}\le\alpha k/m\}\cup\{0\}\right),\qquad
 \cR(\mathbf{p})=\{j:p_j\le\alpha R(\mathbf{p})/m\}.
\end{equation}
When $R(\mathbf{p})=0$, the rejection set is empty. The constructions considered here have positive $p$-values almost surely. Put $V(\mathbf{p})=|\cR(\mathbf{p})\cap \cH_0|$ and
\[
 \FDR(\mathbf{p})=\EE\!\left[\frac{V(\mathbf{p})}{R(\mathbf{p})\vee1}\right].
\]
Unless a conditional probability is explicitly stated, expectations include calibration data, test data, and auxiliary randomness.

The floor $\delta$ implies $R(\bP)=0$ whenever $\delta>\alpha$. More generally, $R(\bP)>0$ requires $R(\bP)\ge\lceil m\delta/\alpha\rceil$. This restriction concerns the number of rejections, not a requirement that every rejected value equal $\delta$.

\begin{Coro}[BH control]\label{cor:bh}
Under the assumptions of Theorem~\ref{thm:transfer}, applying BH at level $0<\alpha<1$ to $\bQ^{\mathrm{tb}}$ gives
\begin{equation}\label{eq:fdrbound}
 \FDR(\bQ^{\mathrm{tb}})\le\pi_0\alpha.
\end{equation}
This bound also holds conditional on the alternative score vector, for almost every realization of that vector.
\end{Coro}

Exactness and PRDS imply the BH guarantee \citep{by2001}; Appendix~\ref{app:bh} gives the argument. A broader formulation separates self-consistency of the rejection rule from conditions on the dependence of the $p$-values \citep{blanchard2008}. This is not a guarantee conditional on the realized calibration sample.

\subsection{Calibration-cell randomization and its limitations}\label{sec:alternatives}
A limitation of tie-block randomization arises when only ordinary $p$-values are retained. Even with the calibration sample known, the same ordinary value can correspond to different tie-block intervals. Thus $(P_j,\bC,U_j)$ need not determine $Q_j^{\mathrm{tb}}$. The missing information is whether the test score equals a calibration score or lies strictly below it.

For example, take $n=1$ and fix $C_1=a$. A score $S<a$ and a score $S=a$ both give $P=1$, but their tie-block randomizations are
\[
 \begin{array}{c|c|c|c}
 \text{Test score}&(A,E)&\text{Tie-block interval}&Q^{\mathrm{tb}}\\ \hline
 S<a&(1,0)&(1/2,1)&(1+U)/2\\
 S=a&(0,1)&(0,1)&U
 \end{array}
\]
For the same draw $U=1/4$, the randomized values are $5/8$ and $1/4$, respectively. This is a deterministic feature of the score-to-rank map and does not require an atomic sampling model. Under independent continuous sampling, however, $S=C_1$ has probability zero, so this particular ambiguity occurs only on a null event.

Atoms make the ambiguity occur with positive probability and can enlarge the difference between intervals. Suppose $C_1=\cdots=C_b=a$, with $n=b$. Both $S<a$ and $S=a$ again give $P=1$, while the respective tie-block intervals are $(b/(b+1),1)$ and $(0,1)$. For a concrete sampling model, let all calibration scores and the null score be i.i.d.\ on $\{a-1,a\}$, with $\PP(S=a)=\rho\in(0,1)$. The two configurations have probabilities $\rho^b(1-\rho)$ and $\rho^{b+1}$. The discrepancy therefore concerns events of positive probability, and the interval widths differ by a factor of $b+1$.

To make the interval depend only on the ordinary value and the calibration sample, define calibration-cell randomization as follows.

For a fixed calibration sample, define the full set of attainable ordinary ranks
\begin{equation}\label{eq:grid}
 \cG(\bC)=\left\{\delta\left(1+\sum_{i=1}^n\ind\{C_i\ge s\}\right):s\in\RR\right\}.
\end{equation}
For $p\in\cG(\bC)$, let $p_{\bC}^-=\max(\{q\in\cG(\bC):q<p\}\cup\{0\})$; thus $P_{j,\bC}^-$ denotes this value at $p=P_j$. With the same independent uniforms as before, define
\begin{equation}\label{eq:cell}
 Q_j^{\mathrm{cell}}=P_{j,\bC}^-+(P_j-P_{j,\bC}^-)U_j.
\end{equation}
The set in (\ref{eq:grid}) ranges over all real score thresholds, including those not observed among the test points. Its cells partition $(0,1)$ up to endpoints. The interval $(P_{j,\bC}^-,P_j)$ is determined by $(P_j,\bC)$, so hypotheses sharing the same ordinary value receive the same interval. In the $n=1$ example, both scores receive $(1/2,1)$, and $Q_j^{\mathrm{cell}}=(1+U_j)/2$. Given $\bC$, the ordinary rank can also be recovered almost surely from the cell containing $Q_j^{\mathrm{cell}}$.

\begin{figure}[!htbp]
\centering
\includegraphics[width=\textwidth]{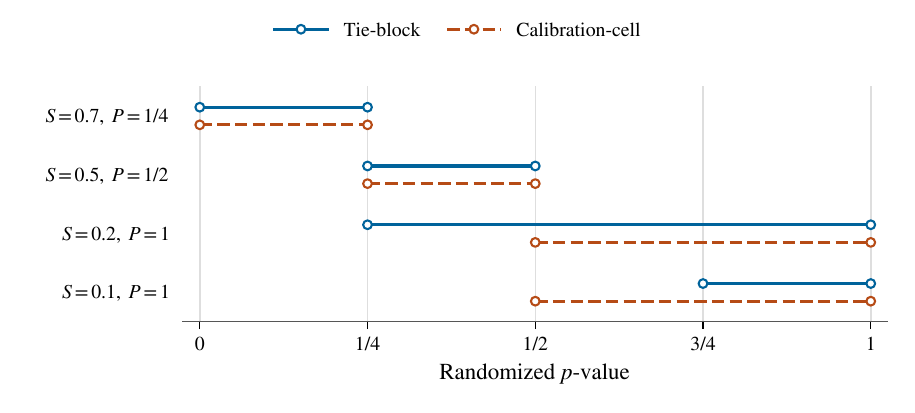}
\caption{Randomization intervals for $n=3$ and $\bC=(.6,.2,.2)$, with attainable ranks $\cG(\bC)=\{1/4,1/2,1\}$. Each row considers a possible test score. Scores $.2$ and $.1$ yield the same ordinary value but different tie-block intervals, whereas their cell intervals coincide. Relative to tie-block smoothing with the same uniform, cell smoothing moves the value upward for $S=.2$ and downward for $S=.1$. Open circles indicate excluded endpoints.}\label{fig:interval-comparison}
\end{figure}

Figure~\ref{fig:interval-comparison} compares the two interval rules for one calibration sample with ties.

This resolves the interval ambiguity, but exact validity of tie-block randomization leaves no room for a uniform improvement toward smaller values. Here a candidate value $\widetilde Q$ is valid when it is super-uniform, that is, when $\PP(\widetilde Q\le t)\le t$ for every $t\in[0,1]$.

\begin{Pro}[No strict downward adjustment of an exact value]\label{prop:no-downward}
Let $Q\sim\Unif(0,1)$. If $\widetilde Q\in[0,1]$ satisfies $\widetilde Q\le Q$ almost surely and $\PP(\widetilde Q<Q)>0$, then $\widetilde Q$ is not a valid $p$-value.
\end{Pro}

Appendix~\ref{app:no-downward} proves this statement. Cell randomization can move values in either direction relative to tie-block randomization, so its validity requires a separate check.

\begin{Pro}[Failure of cell validity]\label{prop:cell}
Let $n=3$, and let the calibration scores and one null test score be independent with common distribution on $\{0,1,2\}$ having probabilities $(9/80,1/10,63/80)$. Then
\begin{equation}\label{eq:cellfail}
 \PP(Q_j^{\mathrm{cell}}\le3/4)=\frac{187573333}{245760000}
 =0.763237846\ldots>\frac34.
\end{equation}
In the same model, $Q_j^{\mathrm{tb}}$ is uniform on $(0,1)$.
\end{Pro}

Appendix~\ref{app:cell} gives the exact enumeration. Recoverability of the ordinary rank does not ensure marginal validity.

Under independent sampling from continuous score distributions, calibration scores are distinct and no test score equals a calibration score almost surely. On this event, $\cG(\bC)=\{\delta,2\delta,\ldots,1\}$, $E_j=0$, and $P_{j,\bC}^-=A_j\delta=P_j-\delta$. Consequently, using the same uniforms,
\begin{equation}\label{eq:ind}
 Q_j=Q_j^{\mathrm{tb}}=Q_j^{\mathrm{cell}}
 =P_j-\delta+\delta U_j=(A_j+U_j)\delta
 \quad\text{almost surely for all }j.
\end{equation}
Thus the ambiguity and the distinction between the two interval rules disappear almost surely in this continuous setting. Continuity of marginal distributions alone is insufficient if scores can be dependent in a way that creates ties with positive probability.

\section{Failure of score-ordered allocation}\label{sec:failure}
\subsection{Ordered constructions and multiset invariance}
Independent tie-block randomization assigns each hypothesis its own uniform draw. Ordered allocation instead sorts the draws within groups and assigns smaller draws to larger scores. The first partition groups by the strict-exceedance count $A_j$, then subdivides each group by the tie count $E_j$; the second groups by the ordinary $p$-value. Thus the groups are
\[
 G_{a,e}^{\mathrm e}=\{j:A_j=a,\ E_j=e\},\qquad
 G_p^{\mathrm p}=\{j:P_j=p\}.
\]
For either type of group $G$, denote the ordered uniforms from $\{U_j:j\in G\}$ by $U_{(1)}(G)<\cdots<U_{(|G|)}(G)$. Order the scores within that group from largest to smallest, resolving score ties by increasing hypothesis index. Let $r_j^{\mathrm e}$ and $r_j^{\mathrm p}$ be the positions under the two partitions, and write
\[
 W_j^{\mathrm e}=U_{(r_j^{\mathrm e})}(G_{A_j,E_j}^{\mathrm e}),\qquad
 W_j^{\mathrm p}=U_{(r_j^{\mathrm p})}(G_{P_j}^{\mathrm p}).
\]
Define the two ordered tie-block rules and the ordered cell rule by
\begin{align}
 Q_j^{\mathrm{tb\text{-}order\text{-}e}}
 &=\delta\{A_j+(E_j+1)W_j^{\mathrm e}\},\label{eq:order-e}\\
 Q_j^{\mathrm{tb\text{-}order\text{-}p}}
 &=\delta\{A_j+(E_j+1)W_j^{\mathrm p}\},\label{eq:order}\\
 Q_j^{\mathrm{cell\text{-}order}}
 &=P_{j,\bC}^-+(P_j-P_{j,\bC}^-)W_j^{\mathrm p}.\label{eq:cell-order}
\end{align}
The first rule sorts uniforms separately within each $(A_j,E_j)$ group, while the other two sort within each ordinary-rank group. Every $G_{a,e}^{\mathrm e}$ is contained in $G_{(a+e+1)\delta}^{\mathrm p}$, so the first partition refines the second. All three rules act within a fixed batch rather than sequentially over time.

\begin{remark}[Ordering within refined groups can reverse score order]\label{rem:order-e}
Let $n=1$, $C_1=1$, $S_1=1$, and $S_2=0$. Both ordinary values equal one, but $(A_1,E_1)=(0,1)$ and $(A_2,E_2)=(1,0)$. The two hypotheses belong to different singleton groups under tb-order-e, so
\[
 Q_1^{\mathrm{tb\text{-}order\text{-}e}}=U_1,\qquad
 Q_2^{\mathrm{tb\text{-}order\text{-}e}}=\frac{1+U_2}{2},
\]
with independent $U_1,U_2$. Whenever $U_1>(1+U_2)/2$, the smaller score $S_2$ receives the smaller randomized value. For example, $U_1=0.9$ and $U_2=0.2$ give values $0.9$ and $0.6$. This reversal occurs on a set of positive probability. Sorting within each $(A_j,E_j)$ group therefore does not ensure score order across groups. The same phenomenon occurs for independent tb randomization in this example, so a reversal alone does not demonstrate invalidity.
\end{remark}

\begin{remark}[Ordering within ordinary-rank groups reinforces separation]\label{rem:order-p}
Within a fixed group $G_p^{\mathrm p}$, a larger score has a weakly smaller $A_j$ and a weakly larger $E_j$, since $A_j+E_j=p/\delta-1$ is fixed. The tb-order-p rule gives that score a smaller assigned uniform as well. Its value can be written as
\begin{equation}\label{eq:order-p-separation}
 Q_j^{\mathrm{tb\text{-}order\text{-}p}}
 =p-\delta(E_j+1)(1-W_j^{\mathrm p}).
\end{equation}
A larger tie count and a smaller assigned uniform both increase the downward displacement from $p$. Thus the allocation reinforces the separation associated with unequal tie-block widths within an ordinary-rank group. In the preceding example, write $U_{(1)}=\min(U_1,U_2)$ and $U_{(2)}=\max(U_1,U_2)$. The ordered values are $U_{(1)}$ and $(1+U_{(2)})/2$, with difference
\[
 \frac{1+U_{(2)}}{2}-U_{(1)}=\frac{1-U_{(1)}}{2}+\frac{U_{(2)}-U_{(1)}}{2}.
\]
The first term is the difference obtained using the common draw $U_{(1)}$ in both intervals; the second is the additional separation from the ordered assignment. This comparison describes the mechanism for fixed draws, rather than asserting that the marginal distributions become more dispersed under every measure of spread. Assigning favorable order statistics according to scores also changes null distributions, as the counterexample below demonstrates even without score ties.
\end{remark}

Under the no-tie conditions in (\ref{eq:ind}), $E_j=0$ and $P_j=(A_j+1)\delta$, so $G_{a,0}^{\mathrm e}=G_{(a+1)\delta}^{\mathrm p}$. The three ordered rules then agree coordinatewise; write $\Qord$ for their common value.

\begin{Pro}[Multiset invariance]\label{prop:multiset}
Suppose calibration scores are distinct and no test score equals a calibration score. Conditional on fixed scores and uniforms, independent smoothing (\ref{eq:ind}), tb-order-e (\ref{eq:order-e}), tb-order-p (\ref{eq:order}), and cell-order (\ref{eq:cell-order}) produce the same multiset of $p$-values. Their BH rejection counts and thresholds are identical.
\end{Pro}

Consequently, under independent sampling from continuous score distributions, if the same original uniforms are used in all constructions,
\[Q_j^{\mathrm{tb}}=Q_j^{\mathrm{cell}}\text{ and }Q_j^{\mathrm{tb\text{-}order\text{-}e}}=Q_j^{\mathrm{tb\text{-}order\text{-}p}}=Q_j^{\mathrm{cell\text{-}order}}
\]
almost surely for every $j$. Write $\bQ$ for the common independent construction and $\Qord$ for the common ordered construction; their sorted values then satisfy $Q_{(k)}=Q_{(k)}^{\mathrm{ord}}$ almost surely for every $k$. Thus all five constructions give the same BH rejection count and threshold almost surely. The equality between independent and ordered constructions concerns their sorted vectors; a given hypothesis can receive different randomized values.

\begin{figure}[!htbp]
\centering
\includegraphics[width=\textwidth]{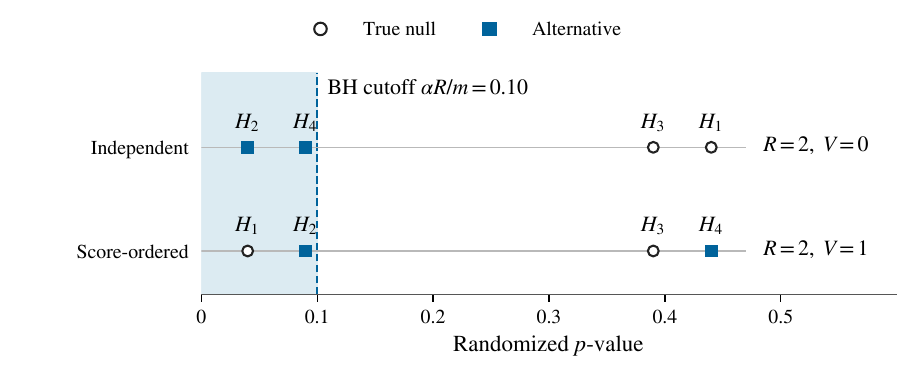}
\caption{A realizable allocation example with $n=1$, $C_1=.5$, $\bS=(.9,.8,.7,.6)$, $\bU=(.88,.08,.78,.18)$, and $\alpha=.2$. The true nulls are $H_1,H_3$ (open circles); alternatives are $H_2,H_4$ (filled squares). All tests occupy the lowest cell. Both allocations produce sorted values $(.04,.09,.39,.44)$, so BH rejects two hypotheses. Shading marks the rejected range; the realized FDP changes from zero to $1/2$. This fixed realization illustrates reassignment, not a marginal FDR calculation.}\label{fig:allocation-comparison}
\end{figure}

Figure~\ref{fig:allocation-comparison} illustrates how the same sorted values can select different hypotheses.

We now restrict attention to continuous scores, for which ordered allocation can already violate both FDR bounds.

\subsection{An exact continuous counterexample}
\begin{Pro}[Two-test failure]\label{prop:minimal}
Let $n=1$, $m=2$, $\alpha=0.05$, and let the calibration score $C$, null score $S_0$, and alternative score $T$ be independent, with $C,S_0\sim\Unif(0,1)$ and
\[
 T\sim\Unif(1/2,1/2+\varepsilon),\qquad 0<\varepsilon<1/2.
\]
Independent smoothing gives $\FDR_{\mathrm{ind}}=0.025$, whereas score-ordered allocation gives
\begin{equation}\label{eq:exacteps}
 \FDR_{\mathrm{ord}}
 =0.030625-0.01125\varepsilon-0.0225\varepsilon^2.
\end{equation}
In particular, the FDR exceeds $\pi_0\alpha=0.025$ for all sufficiently small positive $\varepsilon$. Moreover,
\begin{equation}\label{eq:tail}
 \PP(Q^{\mathrm{ord}}_0\le0.025)
 \longrightarrow0.0309375>0.025
 \quad\text{as }\varepsilon\downarrow0.
\end{equation}
\end{Pro}

Appendix~\ref{app:minimal} gives the finite-$\varepsilon$ calculation. A null with the largest score in its rank group receives the smallest uniform, which can favor a false discovery when only part of the group is rejected. Equation~(\ref{eq:tail}) shows that the resulting null value is not even marginally valid.

The preceding example exceeds $\pi_0\alpha$, but its FDR remains below $\alpha$. In the same model extended to $m/2$ nulls and $m/2$ alternatives, Table~\ref{tab:scale-m} shows that the FDR under score-ordered allocation approaches $\alpha=0.05$ as $m$ increases, while independent smoothing remains close to $\pi_0\alpha=0.025$.

\begin{table}[htbp]
\centering\small
\begin{tabular}{@{}rrr@{}}
\toprule
$m$ & Independent FDR & Score-ordered FDR\\
\midrule
2 & 0.024795 & 0.030155\\
20 & 0.024739 & 0.050503\\
200 & 0.024461 & 0.049310\\
2000 & 0.024893 & 0.049933\\
\bottomrule
\end{tabular}
\caption{Estimated FDR as the number of tests increases, with $n=1$, $\pi_0=1/2$, $\alpha=0.05$, and alternatives distributed as $\Unif(1/2,1/2+10^{-6})$. Each row uses 100000 replications, except $m=2000$, which uses 150000.}\label{tab:scale-m}
\end{table}

\subsection{Failure of the nominal bound and the Simes mechanism}
The preceding observations suggest that score-ordered allocation might still control FDR at $\alpha$ under continuous scores. The rejection count of ordered smoothing agrees with that of independent smoothing, which may suggest a reduction to the complete-null case $\pi_0=1$. Such a reduction is not valid, however, because replacing alternatives by nulls changes the score distribution and hence the distribution of the rejection count. The following example shows that FDR can exceed $\alpha$ even with continuous scores.

The experiment uses $n=19$, $m=1000$, $m_0=500$, and $\alpha=0.05$. Calibration and null scores are i.i.d.\ $\Unif(0,1)$, while the 500 alternatives are independent $\Unif(q,q+10^{-7})$ variables. Thus all scores are continuous.

\begin{table}[htbp]
\centering\small
\begin{tabular}{@{}ccrrrr@{}}
\toprule
$q$ & Allocation & Estimated FDR & $\PP(R>0)$ & $\EE[R]$ & $\EE[R-V]$\\
\midrule
0.97 & Independent & 0.0254 & 0.3115 & 1.57 & 1.44\\
0.97 & Score-ordered & 0.3060 & 0.3115 & 1.57 & 0.14\\
0.99 & Independent & 0.0241 & 0.4308 & 1.98 & 1.86\\
0.99 & Score-ordered & 0.3571 & 0.4308 & 1.98 & 0.76\\
\bottomrule
\end{tabular}
\caption{Paired simulation with 4000 independent replications per alternative location. Both schemes use the same scores and uniforms. The two rightmost columns are Monte Carlo averages. The reference levels are $\pi_0\alpha=0.025$ and $\alpha=0.05$.}\label{tab:ordered}
\end{table}

Table~\ref{tab:ordered} makes the distinction in Proposition~\ref{prop:multiset} visible. Rejection frequencies and mean counts agree, but the number of true discoveries falls under score ordering. The reported FDRs are simulation estimates, not exact calculations. For 4000 independent replications, the worst-case standard error of an average FDP in $[0,1]$ is $1/(2\sqrt{4000})<0.008$; this is an upper bound, not an estimated standard error. The observed excess over $0.05$ is much larger than this scale.

The failure can be understood through the Simes identity \citep{simes1986}. Call the hypotheses sharing one ordinary value a node. The lowest node is $G_\delta^{\mathrm p}$. Conditional on the scores, let $K$ be the number of tests in the lowest node, whose randomized values have the multiset $\{\delta U_1,\ldots,\delta U_K\}$ for independent uniforms. In this experiment $\alpha=\delta=1/(n+1)$, so only this node can be rejected. Write $\beta=\alpha/(m\delta)$ and $\gamma=K\beta$. The Simes identity gives $\PP(R>0\mid \bC,\bS)=\gamma$. Thus an effective level $\gamma$ substantially larger than $\alpha$ makes rejection frequent, although it is certain only at $\gamma=1$. The relevant mismatch is between the node's occupancy $K/m$ and its width $\delta$.

Frequent rejection need not mean many rejections. Appendix~\ref{app:expected-rejections} derives
\[
 \EE[R\mid \bC,\bS]=\sum_{\ell=1}^{K}\ell\,\frac{K!}{(K-\ell)!}\beta^\ell
 \le\frac{\gamma}{(1-\gamma)^2},\qquad 0<\gamma<1.
\]
Hence, when $\gamma$ is bounded away from one, the rejection count remains small relative to $m$, even if $\PP(R>0\mid \bC,\bS)$ is substantial. Here the nonnull scores lie in the fixed narrow interval $(q,q+10^{-7})$. As the number of nulls grows, many null scores exceed all nonnull scores with high probability. Under ordered allocation, these larger null scores receive the smallest randomized $p$-values. If their number exceeds the small rejection count, every rejection is a false discovery. This explains how the FDR can approach the inflated rejection probability.

Let $K_0=\#\{j:S_j>\max_iC_i\}$, and write $\cD=(\bC,\bS)$. The following identity describes why alternatives can increase the opportunity for false discoveries under score ordering.

\begin{Pro}[Lowest-group identity]\label{prop:simes}
Suppose calibration scores are distinct, all calibration--test ties are absent, and $\alpha(n+1)\le1$. Under independent smoothing, tb-order-e, tb-order-p, or cell-order,
\begin{equation}\label{eq:simes}
 \PP(R>0\mid\cD)=\frac{\alpha(n+1)K_0}{m}.
\end{equation}
If the nulls are exchangeable with the calibration sample and the alternatives have a common exceedance probability $p_{\mathrm{alt}}=\PP(S_{\mathrm{alt}}>\max_iC_i)$, then
\begin{equation}\label{eq:gamma}
 \PP(R>0)=\alpha\{\pi_0+(1-\pi_0)(n+1)p_{\mathrm{alt}}\}.
\end{equation}
For score-ordered allocation, $\FDR_{\mathrm{ord}}\le\PP(R>0)$, with equality if and only if every nonempty rejection set contains only nulls almost surely.
\end{Pro}

The proof appears in Appendix~\ref{app:simes}. Equation~(\ref{eq:gamma}) is a rejection probability, not a general formula for FDR. Under the complete null it equals $\alpha$ and also equals FDR. In a mixture, alternatives can increase $K_0$, while score ordering can allocate the smallest uniforms to nulls that are more extreme than those alternatives. This combination explains the much larger FDR in Table~\ref{tab:ordered}.

\section{The ignition process}\label{sec:ignition}
\subsection{BH crossings and rank-cell geometry}
We call $R(\bQ)>R(\bP)$ an \emph{ignition}, including both the first rejection and an increase in an existing rejection set. Throughout this section, condition on $\cD=(\bC,\bS)$, assume distinct calibration scores and no test--calibration ties, and use $Q_j=(A_j+U_j)\delta$. These assumptions hold almost surely under independent continuous sampling. Proposition~\ref{prop:multiset} makes the rejection-count calculations applicable to independent and ordered allocation alike.

\begin{Pro}[Rejection-set inclusion]\label{prop:inclusion}
For any realized data and uniforms in the continuous-score setting,
\[
 \cR(\bP)\subseteq\cR(\bQ).
\]
\end{Pro}

Appendix~\ref{app:inclusion} proves the inclusion for independent smoothing, whose FDR guarantee is given by Corollary~\ref{cor:bh}.

Write $R_0=R(\bP)$, $\widetilde R=R(\bQ)$, and $q_k=\alpha k/m$. Since the data are given, $R_0$ is deterministic. Ignition occurs exactly when a rank beyond $R_0$ satisfies the BH inequality,
\begin{equation}\label{eq:ignition-event}
 \{\widetilde R>R_0\}
 =\bigcup_{k=R_0+1}^{m}\{Q_{(k)}\le q_k\}.
\end{equation}
Because BH is a step-up procedure, it selects the largest crossing rank and rejects the preceding prefix even if earlier ranks fail their own thresholds. The crossing events are therefore not nested. Here the process is a fixed batch, not a sequence of observations over time.

Define, for $b\in\{1,\ldots,n+1\}$,
\[
 G_b=\{j:P_j=b\delta\},\qquad N_b=|G_b|,\qquad
 M_b=\sum_{a<b}N_a.
\]
Node $b$ is $G_b$, the same set as $G_{b\delta}^{\mathrm p}$. Recall that $\beta=\alpha/(m\delta)$.
Values from node $b$ are randomized into $((b-1)\delta,b\delta)$. The intervals of different nodes do not overlap, so node $b$ occupies ranks $M_b+1,\ldots,M_b+N_b$. Let $U_{(r)}(G_b)$ denote its $r$th ordered uniform and put
\begin{equation}\label{eq:node-trigger}
 Q_{(M_b+r)}=(b-1+U_{(r)}(G_b))\delta,\qquad
 c_{b,r}=(M_b+r)\beta-(b-1).
\end{equation}
A crossing at that rank occurs exactly when $U_{(r)}(G_b)\le c_{b,r}$. The node's crossing events depend only on its own uniforms and the fixed group sizes.

BH takes the largest crossing rank and rejects the entire prefix ending there. Thus a crossing in a higher node can bring lower nodes into the rejection set even if they have no crossing themselves. Changing uniforms in a lower node cannot change a higher node's crossing condition. To locate $\widetilde R$, one can scan nonempty nodes in decreasing order of $b$, stop at the first node with a crossing, and take its largest crossing rank. This requires no repeated propagation or re-evaluation of lower nodes. Nodes with $(b-1)\delta\ge\alpha$ can be skipped because their values exceed every BH threshold.

\subsection{Conditional ignition probabilities and the rank floor}
Exact distributions of rejection counts and false discovery proportions for step-up procedures have been studied by \citet{roquain2011}. Here we condition on the scores and use the disjoint rank cells to factor the probabilities over auxiliary uniforms.

For an integer $r\in\{0,\ldots,m\}$, define the event that node $b$ has no crossing above rank $r$,
\[
 \cE_b(r)=\bigcap_{\substack{1\le \ell\le N_b\\M_b+\ell>r}}
 \{U_{(\ell)}(G_b)>c_{b,\ell}\}.
\]
An empty intersection has probability one. A constraint with $c_{b,\ell}\le0$ is automatic, while one with $c_{b,\ell}\ge1$ is impossible almost surely.

\begin{Pro}[Conditional law of the rejection count]\label{prop:ignition-law}
Under the conditions of this section,
\begin{align}
 \PP(\widetilde R\le r\mid\cD)
 &=\prod_{b=1}^{n+1}\PP_{\bU}(\cE_b(r)),\qquad r=0,\ldots,m,\label{eq:ignition-cdf}\\
 \PP(\widetilde R>R_0\mid\cD)
 &=1-\prod_{b=1}^{n+1}\PP_{\bU}(\cE_b(R_0)).\label{eq:ignition-prob}
\end{align}
\end{Pro}

Appendix~\ref{app:nodes} proves the factorization. It gives the entire conditional distribution of $\widetilde R$, not just the probability of obtaining any rejection. Each factor is a probability for uniform order statistics within one node, with affine thresholds $c_{b,\ell}$. The uniforms in different nodes are independent conditional on the scores. Not even independence of the original test scores is required here.

Each factor also admits an exact finite sum based on the independent-uniform Simes identity. Fix a nonempty node, write $N=N_b$, $c_k=c_{b,k}$, and set $s=\max\{1,r-M_b+1\}$. If $s>N$ or $c_N\le0$, then $\PP_{\bU}(\cE_b(r))=1$; if $s\le N$ and $c_N\ge1$, this probability is zero. In the remaining case $s\le N$ and $0<c_N<1$,
\begin{equation}\label{eq:node-simes-sum}
 \PP_{\bU}(\cE_b(r))
 =1-c_N^N
 -(1-c_N)\sum_{\substack{s\le k<N\\c_k>0}}
   \binom Nk c_k^k(1-c_k)^{N-k-1}.
\end{equation}
Indeed, the subtracted terms are the probabilities that the last crossing within the node occurs at rank $N$ or at rank $k<N$, respectively. After conditioning on the $k$ uniforms below $c_k$, the remaining thresholds rescale to $h\beta/(1-c_k)$, so the Simes identity gives their probability of no further crossing. Appendix~\ref{app:nodes} supplies the derivation. Thus (\ref{eq:ignition-cdf})--(\ref{eq:ignition-prob}) can be evaluated directly from the cell occupancies, without simulation.

If $\alpha<\delta$, ordinary BH cannot reject and only node 1 can produce a crossing. Here $N_1=K_0$, in the notation of Proposition~\ref{prop:simes}, and the thresholds are $c_{1,r}=r\beta$. That proposition yields
\begin{equation}\label{eq:ignition-floor}
 \PP(\widetilde R>R_0\mid\cD)
 =N_1\beta=\frac{\alpha(n+1)N_1}{m}.
\end{equation}
This Simes probability accounts for crossings at every rank. Considering only the first threshold gives $1-(1-\beta)^{N_1}$, which underestimates it when $N_1>1$.

At $\alpha=\delta$, the ignition probability is $N_1/m$ if $N_1<m$, and zero if $N_1=m$, since ordinary BH already rejects all hypotheses in the latter case.

Under the complete null and exchangeability, with no ties, $\EE[N_1]=m\delta$, so (\ref{eq:ignition-floor}) gives $\PP(\widetilde R>0)=\FDR=\alpha$ when $\alpha<\delta$. With alternatives present, ignition probability alone measures neither power nor FDR.

\subsection{Activation of higher nodes}
Suppose a nonempty node $b$ lies beyond the ordinary rejection frontier, so $M_b\ge R_0$. Its own randomization can produce a crossing with positive probability exactly when the BH threshold at its last rank enters its interval.

\begin{Pro}[Activation of an unrejected node]\label{prop:node-activation}
For a nonempty node $b$ with $M_b\ge R_0$, a crossing within that node has positive conditional probability if and only if
\begin{equation}\label{eq:node-window}
 (b-1)\delta<q_{M_b+N_b}<b\delta.
\end{equation}
A crossing in this node guarantees ignition. Rejection of its hypotheses can also result from a crossing in a higher node, without the node itself satisfying (\ref{eq:node-window}).
\end{Pro}

The proof is in Appendix~\ref{app:ignition-window}. For an exact example, fix $n=3$, $m=10$, $\alpha=0.6$, with nine ordinary values equal to $0.5$ and one equal to $0.75$. Such ranks are realized by distinct test scores with nine between the largest and second-largest calibration scores and one between the second-largest and smallest. Ordinary BH gives $R_0=9$, since $0.5\le q_9=0.54$ but $0.75>q_{10}=0.6$. The last randomized value is $0.5+0.25U$. Thus $\widetilde R=10$ exactly when $U\le0.4$, and otherwise $\widetilde R=9$. The rejection frontier moves from node 2 to node 3 with probability $0.4$. Appendix~\ref{app:ignition-window} verifies the calculation.

\begin{Pro}[A necessary proximity condition]\label{prop:ignition-near}
If ignition occurs, then for some $k>R_0$,
\begin{equation}\label{eq:ignition-near}
 \#\{j:P_j<q_k+\delta\}\ge k.
\end{equation}
\end{Pro}

Appendix~\ref{app:ignition-window} proves this condition. The condition identifies ranks at which the ordinary values are close enough to a BH threshold for a displacement of less than one grid step to matter. It does not guarantee ignition for a particular realization of the uniforms.

With score ties, tie-block intervals from different nodes can overlap, so nodes need not occupy fixed rank blocks. The node factorization therefore requires the no-tie conditions, whereas the FDR theorem in Section~\ref{sec:fdr} also covers atomic scores.

\section{What smoothing can and cannot repair}\label{sec:limits}
\subsection{The empirical distribution changes within cells}
The empirical distribution of $p$-values determines the BH crossing threshold; its asymptotic role is studied by \citet{genovese2002}. We examine the finite-sample change induced by smoothing the same ordinary ranks.

For a fixed collection of scores with no test--calibration ties, define
\[
 \widehat F_{\bP}(t)=\frac1m\sum_{j=1}^m\ind\{P_j\le t\},\qquad
 \widehat F_{\bQ}(t)=\frac1m\sum_{j=1}^m\ind\{Q_j\le t\}.
\]
The comparison uses the same test points and calibration sample.

\begin{Pro}[ECDF geometry]\label{prop:ecdf}
For independent smoothing, almost surely,
\begin{align}
 \widehat F_{\bP}(t)&\le\widehat F_{\bQ}(t)\le\widehat F_{\bP}(t+\delta),\label{eq:sandwich}\\
 \widehat F_{\bQ}(b\delta)&=\widehat F_{\bP}(b\delta),\quad b=0,\ldots,n+1,\label{eq:grididentity}\\
 \sup_t|\widehat F_{\bQ}(t)-\widehat F_{\bP}(t)|&\le\max_bN_b/m.\label{eq:vertical}
\end{align}
For $t\in[(b-1)\delta,b\delta]$, put $\lambda=(t-(b-1)\delta)/\delta$. Then
\begin{align}
 \EE_{\bU}[\widehat F_{\bQ}(t)\mid\cD]&=\widehat F_{\bP}((b-1)\delta)+\lambda N_b/m,\label{eq:interp}\\
 \Var_{\bU}(\widehat F_{\bQ}(t)\mid\cD)&=N_b\lambda(1-\lambda)/m^2,\label{eq:ecdfvar}
\end{align}
\end{Pro}

Appendix~\ref{app:ecdf} proves these identities. Smoothing preserves the number of points in every rank cell. On average, it replaces each jump by linear interpolation across that cell. Repeating the auxiliary draws can average out the variance in (\ref{eq:ecdfvar}), but it does not change the cell occupancies.

The displacement bound $\delta$ is horizontal. A heavily occupied cell can still produce a large vertical ECDF difference, particularly near zero.

\begin{figure}[!htbp]
\centering
\includegraphics[width=\textwidth]{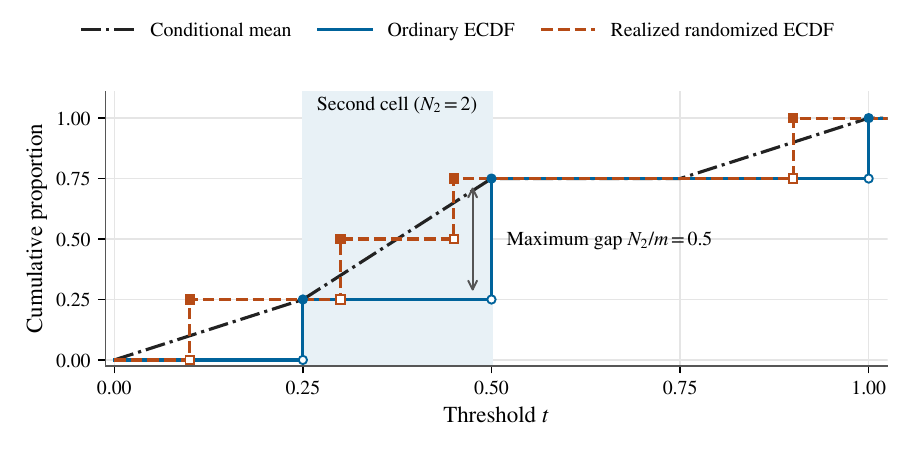}
\caption{Cell occupancy limits the vertical change in the ECDF. With $n=3$, $m=4$, and $\bP=(.25,.5,.5,1)$, the uniforms $\bU=(.4,.2,.8,.6)$ give $\bQ=(.1,.3,.45,.9)$. Shading marks the second rank cell, which contains two observations. On $.45\le t<.5$, the realized ECDF gap is $.75-.25=.5=N_2/m$, attaining the bound in (\ref{eq:vertical}). Filled markers give ECDF values at jumps; open markers give their left limits. Only the conditional mean is linear within cells. All three curves agree at $t=0,.25,.5,.75,1$.}\label{fig:ecdf-schematic}
\end{figure}

Figure~\ref{fig:ecdf-schematic} distinguishes a realized randomized ECDF from its conditional expectation.

\subsection{A calibration-error calculation}
Let $C_1,\ldots,C_n$ be i.i.d.\ calibration scores with continuous null CDF $F_0$, independent of the test score $S$. For a fixed realization $S=s$, let $q(s)=1-F_0(s)$ be the oracle tail probability. Conditional on $S=s$, the number $A$ of calibration scores exceeding $s$ is binomial with parameters $n$ and $q(s)$.

\begin{Pro}[Conditional moments]\label{prop:moments}
For $Q=(A+U_j)/(n+1)$, with independent $U_j\sim\Unif(0,1)$,
\begin{align}
 \EE[Q\mid S=s]&=\frac{nq(s)+1/2}{n+1},\label{eq:mean}\\
 \Var(Q\mid S=s)&=\frac{nq(s)\{1-q(s)\}+1/12}{(n+1)^2},\label{eq:variance}\\
 \EE[(Q-q(s))^2\mid S=s]&=\frac{nq(s)\{1-q(s)\}+1/12+\{1/2-q(s)\}^2}{(n+1)^2}.\label{eq:mse}
\end{align}
\end{Pro}

The proof is in Appendix~\ref{app:moments}. The calibration contribution $nq(s)\{1-q(s)\}/(n+1)^2$ remains after smoothing. At a fixed interior tail probability, its standard deviation is of order $n^{-1/2}$, whereas smoothing moves $P$ by at most $\delta$, of order $n^{-1}$. This comparison is not uniform near the boundary. In particular, thresholds relevant to BH may be much smaller than a fixed interior probability.

Equation~(\ref{eq:mse}) describes approximation to an oracle value conditional on the test score. It is consistent with exact null uniformity after averaging over both the test score and the calibration sample. These are different conditioning statements. It also does not establish a uniform mean-squared-error ordering between ordinary and smoothed values.

The ignition results and these identities describe complementary effects. Smoothing can create discoveries, but it retains the cell occupancies and calibration uncertainty.

Nevertheless, uniform randomization may leave a substantial part of the power loss unrepaired. For comparison, the ordinary value $P=(A+1)/(n+1)$ satisfies
\begin{equation}\label{eq:ordinary-mse}
 \EE[(P-q(s))^2\mid S=s]
 =\frac{nq(s)\{1-q(s)\}+\{1-q(s)\}^2}{(n+1)^2}.
\end{equation}
Equations~(\ref{eq:mse}) and (\ref{eq:ordinary-mse}) share the same leading calibration-error term; smoothing changes only the remaining terms. Figure~\ref{fig:ecdf} illustrates that smoothing can improve rejection frequencies while leaving a gap from the oracle when calibration is limited.

\section{Numerical illustrations}\label{sec:numerics}
\subsection{Normal-model design and rejection frequencies}
We use normal-model experiments to separate the rank-grid effect from the information available to the oracle. Calibration scores follow $N(0,1)$. Test scores follow either $N(0,1)$ under the null or $N(1.5,1)$ under the alternative. The score is the observation itself. The three statistics are the oracle $p_{\mathrm{or}}=1-\Phi(S)$, the ordinary conformal value $P$, and the independently smoothed value $Q$.

For each $n\in\{50,500\}$, one calibration sample is held fixed and $200000$ test observations are generated under each hypothesis. These experiments estimate distributions conditional on the particular calibration sample. Under the alternative, the oracle CDF is
\begin{equation}\label{eq:oraclecdf}
 F_{\mathrm{or},1}(t)=\Phi\{\Phi^{-1}(t)+1.5\},\quad 0<t<1,
\end{equation}
and the null oracle CDF is $t$. Appendix~\ref{app:conditional} gives the corresponding calibration-conditional conformal CDFs.

\begin{table}[htbp]
\centering\small
\begin{tabular}{@{}llrrrr@{}}
\toprule
&&\multicolumn{2}{c}{$n=50$}&\multicolumn{2}{c}{$n=500$}\\
\cmidrule(lr){3-4}\cmidrule(lr){5-6}
Hypothesis & Statistic & Mean & $\PP(p\le.05)$ & Mean & $\PP(p\le.05)$\\
\midrule
Null & Oracle & .4999 & .0507 & .5000 & .0507\\
& Ordinary & .5294 & .0347 & .5041 & .0573\\
& Smoothed & .5196 & .0463 & .5031 & .0573\\
Alternative & Oracle & .1447 & .4410 & .1447 & .4410\\
& Ordinary & .1619 & .3706 & .1418 & .4649\\
& Smoothed & .1520 & .4191 & .1408 & .4652\\
\bottomrule
\end{tabular}
\caption{Normal-model experiments, with one fixed calibration sample per $n$ and $200000$ test draws per hypothesis. Entries are empirical summaries. The last columns give single-test rejection frequencies at a fixed threshold, not BH FDR or multiple-testing power.}\label{tab:normal}
\end{table}

At $n=50$, smoothing increases the alternative rejection frequency from $0.3706$ to $0.4191$, compared with $0.4410$ for the oracle. This is a substantial improvement. At $n=500$, the increase at this threshold is much smaller. These particular values depend on the realized calibration samples. Indeed, at $n=500$ the conditional conformal alternative rejection rate is higher than the oracle's, and the conditional null rejection rate is also higher than $0.05$. This is not a contradiction of marginal conformal validity, nor evidence of a uniformly superior test.

\begin{figure}[!htbp]
\centering
\includegraphics[width=\textwidth]{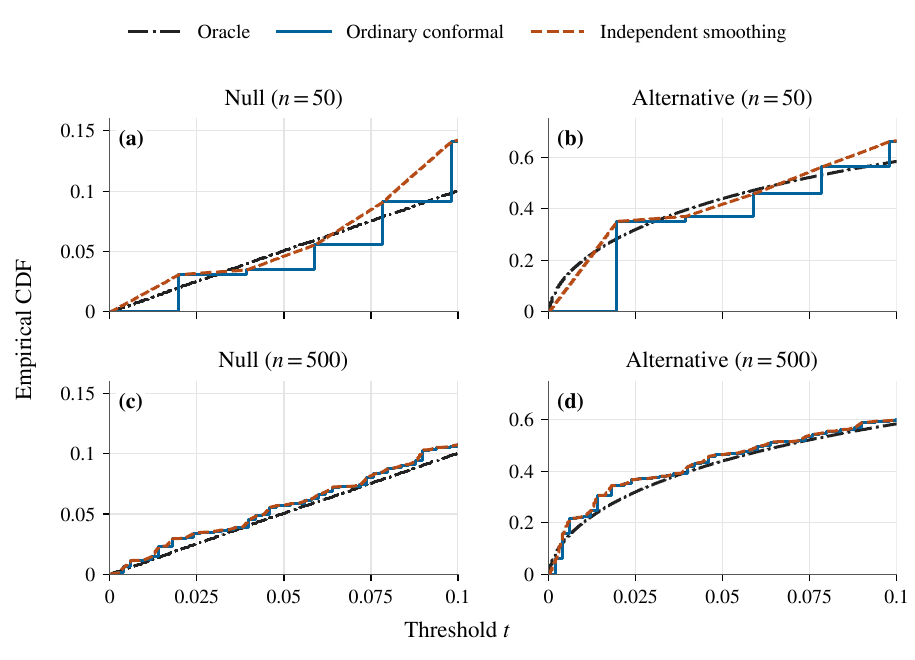}
\caption{Empirical CDFs from the fixed-calibration experiments. The oracle and smoothed curves are shown as steps on the saved threshold grid, spaced by $0.0005$. For ordinary conformal values, jumps are placed at the exact rank thresholds and their heights are read from the saved grid. The panels show thresholds up to $0.10$. Vertical scales differ between null and alternative panels.}\label{fig:ecdf}
\end{figure}

Appendix~\ref{app:normal-experiments} provides additional numerical results, including repeated calibration samples and pooled distribution comparisons.

\subsection{ECDF geometry and calibration variability}
Figure~\ref{fig:ecdf} shows that smoothing changes the shape within the ordinary rank cells. At $n=50$, substantial alternative mass lies in the lowest cell. The ordinary CDF remains zero below $1/51$, while the smoothed CDF can already be large. Across the stored threshold grid over $[0,1]$, the largest ordinary-versus-smoothed differences are approximately $0.0786$ under the null and $0.3494$ under the alternative for $n=50$. For $n=500$, they are $0.0110$ and $0.0736$, respectively. These are maxima over saved thresholds, not exact suprema over all real thresholds.

The practical effect therefore depends on whether the rejection threshold cuts through a heavily occupied cell, as quantified in Proposition~\ref{prop:ecdf}.

The empirical comparisons above condition on one calibration sample. Averaging over fresh calibration samples yields the marginal law, under which null $Q$ is exactly uniform and null $P$ is uniform on its discrete grid. A well-calibrated pooled curve does not imply that each conditional curve lies close to the diagonal. This distinction is formalized in the study of conditional validity for inductive conformal predictors \citep{vovk2012}; see also \citep[Section~4.2.2]{vovk2022}.

The ECDF study, including the additional experiments in Appendix~\ref{app:normal-experiments}, does not directly estimate BH power across calibration samples. Table~\ref{tab:ordered} addresses FDR for a different model, while the normal-model experiments address single-test behavior. We therefore do not infer a general quantitative BH power gain from these plots. A comparison across calibration sizes, signal distributions, and null proportions is needed for that purpose.

\section{Discussion}\label{sec:discussion}
The validity of randomized conformal testing depends on both interval construction and allocation. Independent tie-block smoothing provides a valid route to BH FDR control under the stated exchangeability condition, including with atoms. Neither a common interval for equal ordinary ranks nor preservation of the sorted randomized vector is sufficient. The cell counterexample concerns marginal validity, while ordered allocation changes which hypotheses receive favorable values.

For small calibration samples, the ignition calculation identifies when filling the rank grid can affect decisions. The ECDF and moment identities explain why that improvement can coexist with substantial calibration error. The numerical results support incomplete recovery of oracle single-test power in the normal model, while leaving the size of BH power gains across models unresolved.

Two directions remain open. The first is to weaken the joint exchangeability condition used for tie-block PRDS. The second is to quantify BH power and stability across calibration sizes, signal distributions, and null proportions. Repeating auxiliary draws can assess sensitivity to randomization, but adds no calibration observations; combining such runs would require a separate validity analysis.


\section*{Acknowledgements}
This work was supported by the National Key R\&D Program of China (2022YFA1004801).

\clearpage
\appendix
\setcounter{equation}{0}\renewcommand{\theequation}{A.\arabic{equation}}
\setcounter{figure}{0}\renewcommand{\thefigure}{A.\arabic{figure}}
\setcounter{table}{0}\renewcommand{\thetable}{A.\arabic{table}}
\section{Marginal validity and the cell counterexample}\label{app:tb}
\subsection{Exactness and strict downward adjustments}\paragraph{Proof of Proposition~\ref{prop:tb}.}
\begin{proof}
Fix $j$ as in the statement and put $N=n+1$. Write $Z_r=C_r$ for $1\le r\le n$ and $Z_N=S_j$. Let $\mathcal M$ be the $\sigma$-field generated by the order statistics of $(Z_1,\ldots,Z_N)$, so that conditioning on $\mathcal M$ fixes the pooled multiset, including its multiplicities. Denote its distinct levels by $v_1>\cdots>v_L$, and set
\[
 g_\ell=\sum_{r=1}^N\ind\{Z_r=v_\ell\},\qquad
 B_0=0,\qquad B_\ell=\sum_{h=1}^{\ell}g_h,
 \quad 1\le\ell\le L.
\]
These quantities are $\mathcal M$-measurable, and $B_L=N$. All conditional identities below hold almost surely.

Since $(Z_1,\ldots,Z_N)$ is exchangeable and $\mathcal M$ is invariant under coordinate permutations, the conditional expectations $\EE[\ind\{Z_r=v_\ell\}\mid\mathcal M]$ are equal for all $r$. Hence, for each $1\le\ell\le L$,
\begin{align*}
 \PP(S_j=v_\ell\mid\mathcal M)
 &=\frac1N\sum_{r=1}^N
   \EE[\ind\{Z_r=v_\ell\}\mid\mathcal M]\\
 &=\frac1N\EE\!\left[\sum_{r=1}^N\ind\{Z_r=v_\ell\}
   \,\middle|\,\mathcal M\right]
 =\frac{g_\ell}{N}.
\end{align*}
On the event $\{S_j=v_\ell\}$, the definitions in (\ref{eq:ranks}) give
\[
 A_j=B_{\ell-1},\qquad E_j=g_\ell-1,\qquad
 Q_j^{\mathrm{tb}}=\frac{B_{\ell-1}+g_\ell U_j}{N}.
\]
Independence of $U_j$ from the scores implies that its conditional law given $\mathcal M$ and $S_j$ remains $\Unif(0,1)$. Therefore, conditional on $\mathcal M$ and $S_j=v_\ell$, the law of $Q_j^{\mathrm{tb}}$ is uniform on
\[
 I_\ell=\left(\frac{B_{\ell-1}}N,\frac{B_\ell}N\right),
 \qquad |I_\ell|=\frac{g_\ell}{N}.
\]
Let $\lambda$ denote Lebesgue measure. For any Borel set $I\subseteq[0,1]$, conditioning on the block containing $S_j$ yields
\begin{align*}
 \PP(Q_j^{\mathrm{tb}}\in I\mid\mathcal M)
 &=\sum_{\ell=1}^L\frac{g_\ell}{N}
   \frac{\lambda(I\cap I_\ell)}{g_\ell/N}\\
 &=\sum_{\ell=1}^L\lambda(I\cap I_\ell)
 =\lambda(I).
\end{align*}
The last equality holds because the intervals $I_\ell$ are disjoint and cover $[0,1]$ except for finitely many endpoints, which have Lebesgue measure zero. Taking expectations gives $\PP(Q_j^{\mathrm{tb}}\in I)=\lambda(I)$, proving $Q_j^{\mathrm{tb}}\sim\Unif(0,1)$.
\end{proof}

\paragraph{Proof of Proposition~\ref{prop:no-downward}.}\label{app:no-downward}
\begin{proof}
The event $\{\widetilde Q<Q\}$ is the union, over rational $t\in(0,1)$, of $\{\widetilde Q\le t<Q\}$. Positive probability of the former implies that one of the latter events has positive probability. Since $\widetilde Q\le Q$ almost surely, for this $t$,
\[
 \PP(\widetilde Q\le t)
 =\PP(Q\le t)+\PP(\widetilde Q\le t<Q)>t.
\]
This violates super-uniformity.
\end{proof}

\subsection{Exact enumeration for Proposition~\ref{prop:cell}}\label{app:cell}
\begin{proof}
Let $\mathbf{h}=(h_0,h_1,h_2)=(9/80,1/10,63/80)$. For a calibration configuration $\mathbf{c}$ and test value $x$, denote the cell interval by $(l(\mathbf{c},x),u(\mathbf{c},x))$, as defined in (\ref{eq:grid})--(\ref{eq:cell}). Its conditional CDF at $t$ is
\[
 \varphi(t;l,u)=\min\{1,\max\{0,(t-l)/(u-l)\}\}.
\]
Aggregate the 27 ordered calibration configurations by sorted multiset. Their probabilities $w(\mathbf{c})$ include multinomial multiplicities. The conditional null CDF is $f_{\mathbf{c}}(t)=\sum_{x=0}^2h_x\varphi(t;l(\mathbf{c},x),u(\mathbf{c},x))$. The exact values needed at $t=3/4$ are
\[
\begin{array}{c|r|r}
\mathbf{c}&w(\mathbf{c})&f_{\mathbf{c}}(3/4)\\ \hline
(0,0,0)&729/512000&77/80\\
(0,0,1)&243/64000&151/160\\
(0,0,2)&15309/512000&151/160\\
(0,1,1)&27/8000&71/80\\
(0,1,2)&1701/32000&71/80\\
(0,2,2)&107163/512000&71/80\\
(1,1,1)&1/1000&223/240\\
(1,1,2)&189/8000&143/160\\
(1,2,2)&11907/64000&63/80\\
(2,2,2)&250047/512000&2/3
\end{array}
\]
Summing $w(\mathbf{c})f_{\mathbf{c}}(3/4)$ gives $187573333/245760000$, exceeding $3/4$ by $3253333/245760000>0$. The tie-block statement follows from Proposition~\ref{prop:tb}.
\end{proof}

\paragraph{Conditional validity for a fixed partition.}
Fix $\bC$, and write $F_{\bC}(p)=\PP(P\le p\mid \bC)$. At a cell endpoint $p\in\cG(\bC)$, cell smoothing satisfies
\[
 \PP(Q_j^{\mathrm{cell}}\le p\mid \bC)=F_{\bC}(p).
\]
Between consecutive endpoints its CDF is affine. Consequently, conditional super-uniformity holds if and only if $F_{\bC}(p)\le p$ at every endpoint. Indeed, the difference between that affine CDF and $t$ is affine on each cell, and is nonpositive throughout exactly when it is nonpositive at both endpoints.

This conditional statement does not imply conditional validity of ordinary conformal ranks. Nor can marginal validity fail if every calibration configuration satisfies the conditional criterion. The counterexample has configurations that violate it. Since different configurations have different endpoints, an endpoint for one configuration can lie strictly inside a cell for another.

\section{PRDS and FDR control with atomic scores}\label{app:transfer}
\subsection{Proof of Theorem~\ref{thm:transfer}}
\begin{proof}
Fix $i\in \cH_0$, put $N=n+1$, and write $\bQ=\bQ^{\mathrm{tb}}$. Let $\mathcal M_i$ be the unordered multiset of $C_1,\ldots,C_n,S_i$, retaining all multiplicities, and define
\[
 \mathcal W=(\bS_{-i},\mathcal M_i).
\]
Condition~(\ref{eq:joint-exchangeability}) implies that $(C_1,\ldots,C_n,S_i)$ is exchangeable conditional on $\bS_{-i}$. Indeed, permutations of these $N$ coordinates leave all other test coordinates fixed. Consequently, given $\mathcal W$, the distinguished test score is a uniformly selected copy from the $N$ elements of $\mathcal M_i$.

List the distinct levels of $\mathcal M_i$ as $v_1>\cdots>v_L$, with multiplicities $g_1,\ldots,g_L$. Set $B_0=0$ and $B_\ell=\sum_{r=1}^{\ell}g_r$. Conditional on $\mathcal W$,
\[
 \PP(S_i=v_\ell\mid \mathcal W)=\frac{g_\ell}{N},\qquad
 Q_i\mid(\mathcal W,S_i=v_\ell)\sim
 \Unif\!\left(\frac{B_{\ell-1}}N,\frac{B_\ell}N\right).
\]
Each mixture weight equals the length of its interval, and the intervals partition $(0,1)$ up to endpoints. It follows that
\begin{equation}\label{eq:conditional-uniform}
 Q_i\mid \mathcal W\sim\Unif(0,1),\qquad Q_i\perp \mathcal W.
\end{equation}

Fix a realization $\mathcal W=\omega$. Outside the grid endpoints $\{k/N:0\le k\le N\}$, a value $Q_i=x$ uniquely determines the block containing $S_i$. Denote its score by $v(x)$. The calibration multiset is then $\mathcal M_i$ with one copy of $v(x)$ removed. Its ordering is irrelevant to every randomized rank. The remaining uniforms $\bU_{-i}$ retain their independent uniform law under this conditioning.

For $x<y$, the order of the block intervals gives $v(x)\ge v(y)$. If these levels agree, the calibration multiset is unchanged. Otherwise, passing from the calibration multiset at $x$ to that at $y$ replaces one copy of the smaller score $v(y)$ by the larger score $v(x)$. Couple all other coordinates using the same draws $U_j=u_j$. For $j\ne i$, write
\[
 Q_j=\frac1N\left[u_j+\sum_{r=1}^n\phi(C_r;S_j,u_j)\right],\qquad
 \phi(c;s,u)=
 \begin{cases}
 0,&c<s,\\
 u,&c=s,\\
 1,&c>s.
 \end{cases}
\]
For fixed $s$ and $u\in[0,1]$, $\phi(c;s,u)$ is nondecreasing in $c$. The replacement therefore increases or preserves every $Q_j$, $j\ne i$, while the distinguished coordinate increases from $x$ to $y$. This constructs a coordinatewise monotone coupling of the conditional vectors.

For any increasing measurable set $D$, this coupling shows that
\[
 h_w(x)=\PP(\bQ\in D\mid Q_i=x,\mathcal W=w)
\]
is nondecreasing in $x$ away from the grid endpoints. By (\ref{eq:conditional-uniform}), the mixing law of $\mathcal W$ does not depend on $x$, so
\[
 \PP(\bQ\in D\mid Q_i=x)=\int h_w(x)\,\PP_{\mathcal W}(dw)
\]
is also nondecreasing. The excluded endpoints form a fixed finite set of zero probability under $Q_i$. Assigning monotone one-sided limits there gives a nondecreasing version of the conditional probability. This proves PRDS.

Finally, fix the alternative score vector. For almost every such realization, (\ref{eq:joint-exchangeability}) supplies the same exchangeability assumption under the conditional law, and all uniforms remain independent of the scores. Repeating the argument under that law proves the conditional statement.
\end{proof}

\subsection{Proof of Corollary~\ref{cor:bh}}\label{app:bh}
\begin{proof}
Write $\bQ=\bQ^{\mathrm{tb}}$. Proposition~\ref{prop:tb} gives exact uniformity of every null coordinate, and Theorem~\ref{thm:transfer} gives PRDS. Recall that $q_k=\alpha k/m$, and put $R=R(\bQ)$. For a null $i$, define $f_k(t)=\PP(R\le k\mid Q_i\le t)$. The event $\{R\le k\}$ is increasing in the $p$-value vector. PRDS implies that its conditional probability given $Q_i=x$ is nondecreasing in $x$. Averaging over $x\le t$ shows that $f_k(t)$ is nondecreasing in $t$.

Uniformity and the BH rejection rule now give
\begin{align*}
 \EE\!\left[\frac{\ind\{i\in\cR(\bQ)\}}{R\vee1}\right]
 &=\sum_{k=1}^m\frac{\PP(Q_i\le q_k,R=k)}{k}\\
 &=\frac{\alpha}{m}\sum_{k=1}^m\PP(R=k\mid Q_i\le q_k)\\
 &=\frac{\alpha}{m}\sum_{k=1}^m\{f_k(q_k)-f_{k-1}(q_k)\}.
\end{align*}
For $k\ge2$, monotonicity gives $f_{k-1}(q_k)\ge f_{k-1}(q_{k-1})$. The last sum is at most $f_m(q_m)-f_0(q_1)\le1$. Summing over the $m_0$ null indices establishes $\FDR(\bQ)\le m_0\alpha/m$.

The same argument applies to the conditional law given the alternative score vector. Condition~(\ref{eq:joint-exchangeability}) gives conditional exactness, and Theorem~\ref{thm:transfer} gives conditional PRDS. Thus the conditional FDR is also at most $\pi_0\alpha$ almost surely.
\end{proof}

\section{Ordered allocation and rejection probabilities}
\subsection{Proof of Proposition~\ref{prop:multiset}}\label{app:multiset}
\begin{proof}
Fix the scores and uniforms. Under the stated no-tie conditions, $E_j=0$ and $P_{j,\bC}^-=A_j\delta$. For $a\in\{0,\ldots,n\}$, put $G_a=\{j:A_j=a\}$. Then
\[
 G_a=G_{a,0}^{\mathrm e}=G_{(a+1)\delta}^{\mathrm p},
\]
so all three ordered rules agree. Sorting and assigning the uniforms within each nonempty $G_a$ defines a permutation $\pi_a:G_a\to G_a$ such that
\[
 Q_j^{\mathrm{ord}}=(a+U_{\pi_a(j)})\delta
 =Q_{\pi_a(j)},\qquad j\in G_a,
\]
where $Q_j=(A_j+U_j)\delta$ is the independently smoothed value. Equal uniforms may be assigned in any order, without changing these values. Since the groups partition $\{1,\ldots,m\}$, the maps $\pi_a$ combine into a permutation of all indices. Thus the two vectors have the same multiset of entries, and $Q_{(k)}^{\mathrm{ord}}=Q_{(k)}$ for every $k$. By (\ref{eq:bh}),
\[
 R(\Qord)
 =\max\left(\{k:Q_{(k)}\le\alpha k/m\}\cup\{0\}\right)
 =R(\bQ),
\]
where $k$ ranges over $\{1,\ldots,m\}$. Their BH thresholds $\alpha R/m$ are consequently identical.
\end{proof}

\subsection{Proof of Proposition~\ref{prop:minimal}}\label{app:minimal}
\begin{proof}
First fix the alternative score at $T=t\in(0,1)$. Only the group with $A=0$ can be rejected, since $\alpha=0.05<1/2$. There are three relevant score events. On $\cE_{\max}=\{C<t<S_0\}$, the null is the larger score in a two-point lowest group. On $\cE_{\min}=\{C<S_0<t\}$, it is the smaller score in that group. On $\cE_{\mathrm{alone}}=\{t<C<S_0\}$, it is alone in the lowest group. Their probabilities are $t(1-t)$, $t^2/2$, and $(1-t)^2/2$, respectively. In all remaining events the null cannot be rejected.

Let $U_{(1)}=\min(U_0,U_1)$ and $U_{(2)}=\max(U_0,U_1)$, and temporarily allow $0<\alpha<1/2$. If both hypotheses are in the lowest group, $R=2$ when $U_{(2)}\le2\alpha$, whereas $R=1$ when $U_{(1)}\le\alpha$ and $U_{(2)}>2\alpha$. The latter event has probability $2\alpha(1-2\alpha)$. Under score ordering, the conditional expected FDP on $\cE_{\max}$ is therefore $2\alpha(1-2\alpha)+(1/2)(2\alpha)^2=2\alpha-2\alpha^2$. On $\cE_{\min}$ it is $2\alpha^2$. On $\cE_{\mathrm{alone}}$, the null is rejected exactly when $U_0\le\alpha$, so the expected FDP is $\alpha$.

Combining the three events yields
\begin{align}
 \FDR_{\mathrm{ord}}(t)
 &=(2\alpha-2\alpha^2)t(1-t)+\alpha^2t^2+\frac{\alpha}{2}(1-t)^2\notag\\
 &=\frac{\alpha}{2}+\alpha(1-2\alpha)\left(t-\frac32t^2\right).\label{eq:ft}
\end{align}
For $T\sim\Unif(1/2,1/2+\varepsilon)$, $\EE[T]=1/2+\varepsilon/2$ and $\EE[T^2]=1/4+\varepsilon/2+\varepsilon^2/3$. At $\alpha=0.05$, averaging (\ref{eq:ft}) gives (\ref{eq:exacteps}).

Under independent allocation, the conditional expected FDP whenever both points are in the lowest group is $\alpha(1-2\alpha)+(1/2)(2\alpha)^2=\alpha$. The probability of that joint event is $t-t^2/2$. Adding the alone event gives
\[
 \FDR_{\mathrm{ind}}(t)=\alpha(t-t^2/2)+\alpha(1-t)^2/2=\alpha/2.
\]
It remains $\alpha/2$ after averaging over $T$.

For the marginal tail, let $t=1/2$ and use the threshold $\alpha/2=0.025$. On $\cE_{\max}$, the relevant probability is $\PP(U_{(1)}\le\alpha)=2\alpha-\alpha^2$; on $\cE_{\min}$, it is $\PP(U_{(2)}\le\alpha)=\alpha^2$; on the alone event, it is $\alpha$. Their weights are $1/4,1/8,1/8$. At $\alpha=0.05$, the sum is $0.0309375$. For the narrow uniform alternative these event probabilities are continuous in $t$, so averaging and sending $\varepsilon$ to zero gives (\ref{eq:tail}).
\end{proof}

\subsection{Simes crossing identities}\label{app:uniform}
The Simes identity \citep{simes1986} states that, for independent uniforms $Z_1,\ldots,Z_K$ and $0\le\gamma\le1$,
\begin{equation}\label{eq:simesproof}
 \PP\left(\exists k\le K:Z_{(k)}\le\gamma k/K\right)=\gamma.
\end{equation}
Here is a direct proof. Write $\bZ=(Z_1,\ldots,Z_K)$. Let $R$ be its BH rejection count at level $\gamma$, and let $R_i^0$ be the count when coordinate $i$ is replaced by zero. For $r\ge1$,
\[
 \{Z_i\le\gamma r/K,\ R=r\}
 =\{Z_i\le\gamma r/K,\ R_i^0=r\}.
\]
If $R=r$ and $Z_i$ is rejected, lowering that already rejected coordinate leaves all order statistics above rank $r$ unchanged. Conversely, if $R_i^0=r$ and the restored value remains below $\gamma r/K$, the $r$th crossing is retained and no larger crossing can be created by increasing a coordinate. Thus the event identity holds. Independence of $R_i^0$ from $Z_i$, and $R_i^0\ge1$, imply
\[
 \EE\!\left[\frac{\ind\{i\text{ is rejected}\}}{R\vee1}\right]
 =\sum_{r=1}^K\frac{(\gamma r/K)\PP(R_i^0=r)}r=\gamma/K.
\]
Summing over all $K$ coordinates gives $\PP(R>0)=\gamma$, since all $K$ hypotheses are null.

\subsection{Proof of Proposition~\ref{prop:simes}}\label{app:simes}
\begin{proof}
The condition $\alpha(n+1)\le1$ is equivalent to $\alpha\le\delta$. Every smoothed value outside the lowest group exceeds $\delta$ almost surely and hence exceeds every BH threshold. Conditional on $\cD$, the $K_0$ values in the lowest group have multiset $\{\delta U_1,\ldots,\delta U_{K_0}\}$. Ordered allocation preserves this multiset. For $K_0>0$, a rejection occurs exactly when
\[
 \exists k\le K_0:\ U_{(k)}\le\frac{\alpha k}{m\delta}
 =\frac{k}{K_0}\frac{\alpha K_0}{m\delta}.
\]
Apply (\ref{eq:simesproof}) with $\gamma=\alpha K_0/(m\delta)\le1$. If $K_0=0$, both sides of (\ref{eq:simes}) are zero.

Each null score exceeds the calibration maximum with probability $\delta$, by exchangeability and absence of ties. Thus $\EE[K_0]=m_0\delta+(m-m_0)p_{\mathrm{alt}}$, giving (\ref{eq:gamma}). Finally, $V/(R\vee1)\le\ind\{R>0\}$ pointwise. Equality of expectations holds exactly when $V/R=1$ on every nonempty rejection event, up to a null set.
\end{proof}

\subsection{The expected number of rejections and extreme null scores}\label{app:expected-rejections}
Assume the no-tie conditions of Proposition~\ref{prop:simes} and $\alpha\le\delta$. Conditional on the scores, set $K=K_0$ and $\gamma=K\beta$, with $\beta=\alpha/(m\delta)$ as before. Only the lowest node can be rejected, and its rejection count is
\[
 R=\max\bigl(\{r\le K:U_{(r)}\le r\beta\}\cup\{0\}\bigr).
\]
If $K=0$, then $R=0$. If $\gamma=1$, necessarily $\alpha=\delta$ and $K=m$, so $R=m$. The following calculation concerns $K\ge1$ and $0<\gamma<1$.

For $0\le r<K$, the event $R=r$ requires exactly $r$ uniforms in $[0,r\beta]$. Conditional on their identities, the remaining $K-r$ uniforms are independent on $(r\beta,1)$. After rescaling that interval to $(0,1)$, the absence of any later crossing has probability
\[
 1-\frac{(K-r)\beta}{1-r\beta}
 =\frac{1-K\beta}{1-r\beta},
\]
by the independent-uniform identity in Appendix~\ref{app:uniform}. Therefore
\begin{equation}\label{eq:rejection-count-law}
 \PP(R=r\mid \bC,\bS)
 =\binom Kr(r\beta)^r(1-r\beta)^{K-r-1}(1-K\beta),
 \qquad 0\le r<K,
\end{equation}
where $0^0=1$. The remaining mass is $\PP(R=K\mid \bC,\bS)=(K\beta)^K$. Multiplying by $r$ and summing gives the exact expectation. Expanding the factors $(1-r\beta)^{K-r-1}$, collecting powers of $\beta$, and including the $r=K$ term yields
\begin{equation}\label{eq:expected-rejections}
 \EE[R\mid \bC,\bS]
 =\sum_{\ell=1}^{K}\ell\,\frac{K!}{(K-\ell)!}\beta^\ell
 \le\sum_{\ell=1}^{\infty}\ell\gamma^\ell
 =\frac{\gamma}{(1-\gamma)^2}.
\end{equation}
The inequality uses $K!/(K-\ell)!\le K^\ell$. Since $\PP(R>0\mid \bC,\bS)=\gamma$, it also gives $\EE[R\mid R>0,\bC,\bS]\le(1-\gamma)^{-2}$. For example, $\gamma=0.3$ permits rejection with probability $0.3$ while bounding the mean count conditional on rejection by $1/0.7^2<2.05$. This example illustrates the bound rather than substituting a fixed effective level for the random one in Table~\ref{tab:ordered}.

To compare this count with extreme null scores, let $s_*=q+10^{-7}<1$, the upper endpoint of the alternative distribution, and put
\[
 L_0=\#\{j\in \cH_0:S_j>\max(s_*,\max_iC_i)\}.
\]
These $L_0$ nulls all belong to the lowest node and have larger scores than every alternative. Under ordered allocation, $R\le L_0$ therefore implies $V=R$. The conditional bound from (\ref{eq:expected-rejections}) is
\[
 \PP(R>L_0\mid \bC,\bS)
 \le\frac{\gamma}{L_0(1-\gamma)^2},\qquad L_0\ge1.
\]
Consequently, the difference between $\PP(R>0\mid \bC,\bS)$ and the conditional FDR is at most this probability.

More precisely, keep $n$ fixed and let $m\to\infty$ with $m_0/m\to\pi_0\in(0,1)$ in the independent sampling model of Table~\ref{tab:ordered}. Conditional on any calibration sample with $C_{\max}:=\max_i C_i<1$, the law of large numbers gives $L_0/m\to\pi_0\{1-\max(s_*,C_{\max})\}>0$ almost surely. In this experiment $\alpha=\delta$, so $\gamma=K/m$, and the same law gives $\limsup\gamma\le1-\pi_0 C_{\max}<1$ almost surely. Thus (\ref{eq:expected-rejections}) bounds the conditional expected rejection count while $L_0$ grows linearly, and $\PP(R>L_0\mid \bC,\bS)\to0$. Calibration scores are uniform, so $0<C_{\max}<1$ almost surely. Averaging the bounded conditional probabilities shows that $\PP(R>0)-\FDR_{\mathrm{ord}}\to0$. The mechanism is that the number of extreme nulls grows while the number of rejected hypotheses remains small.

\section{Smoothing identities and conditional laws}
\subsection{Ignition and node crossings}\paragraph{Proof of Proposition~\ref{prop:inclusion}.}\label{app:inclusion}
\begin{proof}
Coordinatewise $\bQ\le \bP$ implies $Q_{(k)}\le P_{(k)}$ at every rank. Therefore $R(\bQ)\ge R(\bP)$. If $j\in\cR(\bP)$, then
\[
 Q_j\le P_j\le\alpha R(\bP)/m\le\alpha R(\bQ)/m,
\]
so $j\in\cR(\bQ)$.
\end{proof}

\paragraph{Proof of Proposition~\ref{prop:ignition-law}.}\label{app:nodes}
\begin{proof}
Conditional on $\cD$, recall $G_b=\{j:P_j=b\delta\}$, $N_b=|G_b|$, and $M_b=\sum_{a<b}N_a$. The observations in group $b$ occupy exactly ranks $M_b+1,\ldots,M_b+N_b$ after independent smoothing. Write $U_{(1)}(G_b),\ldots,U_{(N_b)}(G_b)$ for the group's ordered uniforms. For a set of ranks $\cN\subseteq\{1,\ldots,m\}$, put
\[
 J_b(\cN)=\{r\in\{1,\ldots,N_b\}:M_b+r\in \cN\},\qquad
 c_{b,r}=\frac{\alpha(M_b+r)}{m\delta}-(b-1).
\]
Then
\begin{equation}\label{eq:nodeprod}
 \PP\left(Q_{(k)}>\alpha k/m\text{ for every }k\in \cN\mid\cD\right)
 =\prod_{b=1}^{n+1}
 \PP\left(U_{(r)}(G_b)>c_{b,r}\text{ for every }r\in J_b(\cN)\right).
\end{equation}
To verify the identity, use $Q_{(M_b+r)}=(b-1+U_{(r)}(G_b))\delta$. The constraints on different groups depend on disjoint sets of independent uniforms, so their probabilities multiply. A constraint with $c_{b,r}\le0$ is automatic; one with $c_{b,r}\ge1$ is impossible. Empty intersections have probability one.

Taking $\cN=\{1,\ldots,m\}$ gives the probability of no rejection. Taking $\cN=\{R(\bP)+1,\ldots,m\}$, together with Proposition~\ref{prop:inclusion}, gives the probability that smoothing produces no additional rejections. No first-threshold approximation is used.

Taking $\cN=\{r+1,\ldots,m\}$ gives $\{\widetilde R\le r\}$, since BH has no crossing above $r$ exactly on this event. The node events in (\ref{eq:nodeprod}) are then $\cE_b(r)$, yielding (\ref{eq:ignition-cdf}). Rejection-set inclusion gives $\widetilde R\ge R_0$, so its specialization to $r=R_0$ yields (\ref{eq:ignition-prob}).
\end{proof}

\begin{proof}[Derivation of (\ref{eq:node-simes-sum})]
Fix the scores and a nonempty node, suppress its index, and write $U_{(1)}\le\cdots\le U_{(N)}$ for its ordered uniforms. Define its last crossing rank by
\[
 L=\max\bigl(\{k\in\{1,\ldots,N\}:U_{(k)}\le c_k\}\cup\{0\}\bigr).
\]
For $s=\max\{1,r-M_b+1\}$, the event $\cE_b(r)$ is $\{L<s\}$. The stated boundary cases follow immediately from the increasing thresholds. Suppose henceforth that $s\le N$ and $0<c_N<1$. For $1\le k<N$ with $c_k>0$, the event $\{L=k\}$ requires exactly $k$ uniforms in $[0,c_k]$. Indeed, fewer would preclude the crossing at $k$, while more would imply $U_{(k+1)}\le c_k<c_{k+1}$ and hence a later crossing.

Conditional on exactly $k$ such uniforms and their identities, the remaining $N-k$ variables are independent and uniform on $(c_k,1)$. Rescale them to independent $\Unif(0,1)$ variables $V_1,\ldots,V_{N-k}$. Since $c_{k+h}-c_k=h\beta$, absence of a later crossing is equivalent to
\[
 V_{(h)}>\frac{h\beta}{1-c_k}
 =\frac{h}{N-k}\gamma_k,\qquad
 1\le h\le N-k,\qquad
 \gamma_k=\frac{(N-k)\beta}{1-c_k}<1.
\]
The inequality follows from $c_N=c_k+(N-k)\beta<1$. By the independent-uniform Simes identity (\ref{eq:simesproof}), the conditional probability of these inequalities is $1-\gamma_k$. Consequently,
\[
 \PP_{\bU}(L=k)
 =\binom Nk c_k^k(1-c_k)^{N-k}(1-\gamma_k)
 =\binom Nk c_k^k(1-c_k)^{N-k-1}(1-c_N).
\]
For $c_k\le0$, $\PP_{\bU}(L=k)=0$, while $\PP_{\bU}(L=N)=c_N^N$. Summing these disjoint last-crossing events for $k\ge s$ and taking the complement proves (\ref{eq:node-simes-sum}).
\end{proof}

\paragraph{Node activation and the proximity condition.}\label{app:ignition-window}
\begin{proof}[Proof of Proposition~\ref{prop:node-activation}]
Write $k_b=M_b+N_b$. Because $k_b>R_0$, the ordinary value $P_{(k_b)}=b\delta$ cannot satisfy the BH inequality, so $q_{k_b}<b\delta$. The thresholds $c_{b,r}$ increase in $r$. If $q_{k_b}\le(b-1)\delta$, they are all nonpositive and there is no crossing in the node. If $q_{k_b}>(b-1)\delta$, then $c_{b,N_b}\in(0,1)$, and the event that every uniform in the node is at most $c_{b,N_b}$ has probability $c_{b,N_b}^{N_b}>0$. On that event rank $k_b$ crosses. Since every rank in the node exceeds $R_0$, any crossing there guarantees ignition. A higher crossing rejects the entire preceding prefix by the BH rule, regardless of the node's own crossings.
\end{proof}

In the example, the first nine randomized values are below $0.5<q_9$, so $\widetilde R\ge9$ for every draw. The remaining value is at rank 10 and crosses exactly when $0.5+0.25U\le0.6$, or $U\le0.4$. This proves the stated two-point conditional law.

\begin{proof}[Proof of Proposition~\ref{prop:ignition-near}]
On ignition, choose a crossing rank $k>R_0$. At least $k$ coordinates satisfy $Q_j\le q_k$. Since $P_j-\delta<Q_j$ almost surely, each such coordinate satisfies $P_j<q_k+\delta$. Counting them proves (\ref{eq:ignition-near}).
\end{proof}

\subsection{ECDF geometry and calibration error}\paragraph{Proof of Proposition~\ref{prop:ecdf}.}\label{app:ecdf}
\begin{proof}
The inequalities $P_j-\delta<Q_j<P_j$ imply $\ind\{P_j\le t\}\le\ind\{Q_j\le t\}\le\ind\{P_j\le t+\delta\}$. Summing proves (\ref{eq:sandwich}). At a grid point $b\delta$, values from lower cells are already below the threshold and values from higher cells are above it. Therefore exactly the same observations contribute to both ECDFs, proving (\ref{eq:grididentity}).

Inside cell $b$, the only possible difference is contributed by its $N_b$ observations. This proves (\ref{eq:vertical}). Conditional on all scores, each of these observations falls below $t$ independently with probability $\lambda$. Thus
\[
 m\widehat F_{\bQ}(t)=m\widehat F_{\bP}((b-1)\delta)+B,\qquad
 B\mid\cD\sim\mathrm{Bin}(N_b,\lambda),
\]
which gives (\ref{eq:interp}) and (\ref{eq:ecdfvar}), including the endpoints.
\end{proof}

\paragraph{Proof of Proposition~\ref{prop:moments}.}\label{app:moments}
\begin{proof}
Conditional on $S=s$, the indicators $\ind\{C_i>s\}$ are independent Bernoulli variables with probability $q(s)$. Hence $\EE[A\mid s]=nq(s)$ and $\Var(A\mid s)=nq(s)\{1-q(s)\}$. Since $U_j$ is independent with mean $1/2$ and variance $1/12$, dividing the moments of $A+U_j$ by $n+1$ gives (\ref{eq:mean}) and (\ref{eq:variance}). The conditional bias relative to $q(s)$ is $\{1/2-q(s)\}/(n+1)$. Adding its square to the variance gives (\ref{eq:mse}).

For $P=(A+1)/(n+1)$, the conditional variance is $nq(s)\{1-q(s)\}/(n+1)^2$ and the conditional bias is $\{1-q(s)\}/(n+1)$. Adding the squared bias to the variance proves (\ref{eq:ordinary-mse}).
\end{proof}

\subsection{Calibration-conditional CDFs}\label{app:conditional}
Suppose the calibration values are distinct. Order them as $C_{[1]}>\cdots>C_{[n]}$, and set $C_{[0]}=+\infty$, $C_{[n+1]}=-\infty$. If a test score $S$ is independent of $\bC$ and has continuous CDF $F_S$, then, conditional on $\bC$,
\[
 w_k(\bC;F_S)=\PP(A=k\mid \bC)=F_S(C_{[k]})-F_S(C_{[k+1]}),\qquad k=0,\ldots,n.
\]
Indeed, $A=k$ is the event $C_{[k+1]}<S<C_{[k]}$, up to zero-probability boundaries. Thus
\begin{align*}
 \PP(P\le t\mid \bC)&=\sum_{k=0}^nw_k(\bC;F_S)\ind\{(k+1)\delta\le t\},\\
 \PP(Q\le t\mid \bC)&=\sum_{k=0}^nw_k(\bC;F_S)
 \min\{1,\max\{0,t/\delta-k\}\}.
\end{align*}
The weights are not generally equal conditional on $\bC$, even when $F_S=F_0$. Under null exchangeability their unconditional expectations equal $\delta$. For the normal oracle, $S\sim N(1.5,1)$ and
\[
 \PP(1-\Phi(S)\le t)
 =\PP(S\ge\Phi^{-1}(1-t))
 =\Phi\{\Phi^{-1}(t)+1.5\},
\]
establishing (\ref{eq:oraclecdf}).

\section{Additional normal-model experiments}\label{app:normal-experiments}
\subsection{Protocol and reference distributions}
We extend the normal-model design of Section~\ref{sec:numerics}, using $n\in\{50,500\}$, null scores $N(0,1)$, and alternative scores $N(1.5,1)$. The oracle CDF is given by (\ref{eq:oraclecdf}); Appendix~\ref{app:conditional} gives the calibration-conditional laws of $P$ and $Q$.

The fixed-calibration experiment uses one calibration sample and $200000$ test observations per hypothesis, as in Section~\ref{sec:numerics}. A second experiment draws $100$ independent calibration samples, with $10000$ tests per hypothesis for each sample. Its pooled ECDF averages the $100$ calibration-conditional ECDFs, giving $10^6$ observations per curve. Tests within a calibration batch share the same calibration sample; the pooled observations therefore do not constitute $10^6$ independent calibration replications. A separate check redraws calibration for each of $4000$ independent single-test replications.

For $0\le t\le1$, the exact marginal null CDFs of $P$ and $Q$ are $\lfloor(n+1)t\rfloor/(n+1)$ and $t$, respectively. These marginal statements do not require the conditional CDFs for one calibration sample to equal their references. Under the alternative, the marginal rank weights are $\EE[w_k(\bC;F_{\mathrm{alt}})]$, with $F_{\mathrm{alt}}(s)=\Phi(s-1.5)$, and are generally unequal. In the saved repeated-calibration experiment, their reference values are obtained by averaging the conditional weights over $40000$ additional calibration draws. References based on these weights are numerical approximations to the marginal laws.

\subsection{Fixed calibration and fresh-calibration checks}
Table~\ref{tab:normal-small-thresholds} supplements Table~\ref{tab:normal} with smaller thresholds. For $n=50$, ordinary conformal testing cannot reject at $t=.001$ or $.01$, since both are below $1/51$. Smoothing creates rejections at these thresholds, but the alternative frequencies remain below the oracle's. At $n=500$, smoothing has a smaller effect at $.01$, whereas its removal of the rank floor remains visible at $.001$. Conditional frequencies can exceed the oracle's at some thresholds; the accompanying null frequencies also depend on the realized calibration sample.

\begin{table}[htbp]
\centering\small
\begin{tabular}{@{}llrrrr@{}}
\toprule
&&\multicolumn{2}{c}{$n=50$}&\multicolumn{2}{c}{$n=500$}\\
\cmidrule(lr){3-4}\cmidrule(lr){5-6}
Hypothesis & Statistic & $t=.001$ & $t=.01$ & $t=.001$ & $t=.01$\\
\midrule
Null & Oracle & .0010 & .0099 & .0010 & .0099\\
& Ordinary & .0000 & .0000 & .0000 & .0123\\
& Smoothed & .0014 & .0155 & .0006 & .0123\\
Alternative & Oracle & .0559 & .2034 & .0559 & .2036\\
& Ordinary & .0000 & .0000 & .0000 & .2253\\
& Smoothed & .0185 & .1797 & .0299 & .2255\\
\bottomrule
\end{tabular}
\caption{Empirical rejection frequencies $\PP(p\le t\mid\bC)$ from the fixed-calibration experiments, with $200000$ test draws per hypothesis.}\label{tab:normal-small-thresholds}
\end{table}

Table~\ref{tab:normal-fresh} reports selected results from the separate fresh-calibration check. At $n=50$ and $t=.01$, the alternative frequency increases from zero for $P$ to $.1245$ for $Q$, compared with $.2028$ for the oracle. At $t=.05$, the corresponding values are $.3735$, $.4190$, and $.4437$. Thus the incomplete recovery also appears when averaging over calibration samples, rather than only for the fixed sample in Figure~\ref{fig:ecdf}. These are single-test frequencies, not BH power estimates.

\begin{table}[htbp]
\centering\small
\begin{tabular}{@{}rr rrr rrr@{}}
\toprule
&&\multicolumn{3}{c}{Null}&\multicolumn{3}{c}{Alternative}\\
\cmidrule(lr){3-5}\cmidrule(lr){6-8}
$n$&$t$&Oracle&Ordinary&Smoothed&Oracle&Ordinary&Smoothed\\
\midrule
50 & .001 & .0015 & .0000 & .0015 & .0580 & .0000 & .0118\\
& .010 & .0135 & .0000 & .0130 & .2028 & .0000 & .1245\\
& .050 & .0540 & .0455 & .0568 & .4437 & .3735 & .4190\\
& .100 & .1015 & .1030 & .1052 & .5737 & .5563 & .5605\\
500 & .001 & .0018 & .0000 & .0013 & .0617 & .0000 & .0395\\
& .010 & .0140 & .0127 & .0127 & .2087 & .2003 & .2003\\
& .050 & .0517 & .0515 & .0515 & .4402 & .4373 & .4375\\
& .100 & .0975 & .0993 & .0993 & .5847 & .5787 & .5790\\
\bottomrule
\end{tabular}
\caption{Empirical rejection frequencies with fresh calibration in each of $4000$ independent replications per hypothesis and calibration size. For any one entry, the binomial standard error is at most $1/(2\sqrt{4000})\approx.0079$.}\label{tab:normal-fresh}
\end{table}

\subsection{Calibration variability, pooled fits, and oracle comparisons}
\begin{figure}[p]
\centering
\includegraphics[width=.94\textwidth,height=.76\textheight,keepaspectratio]{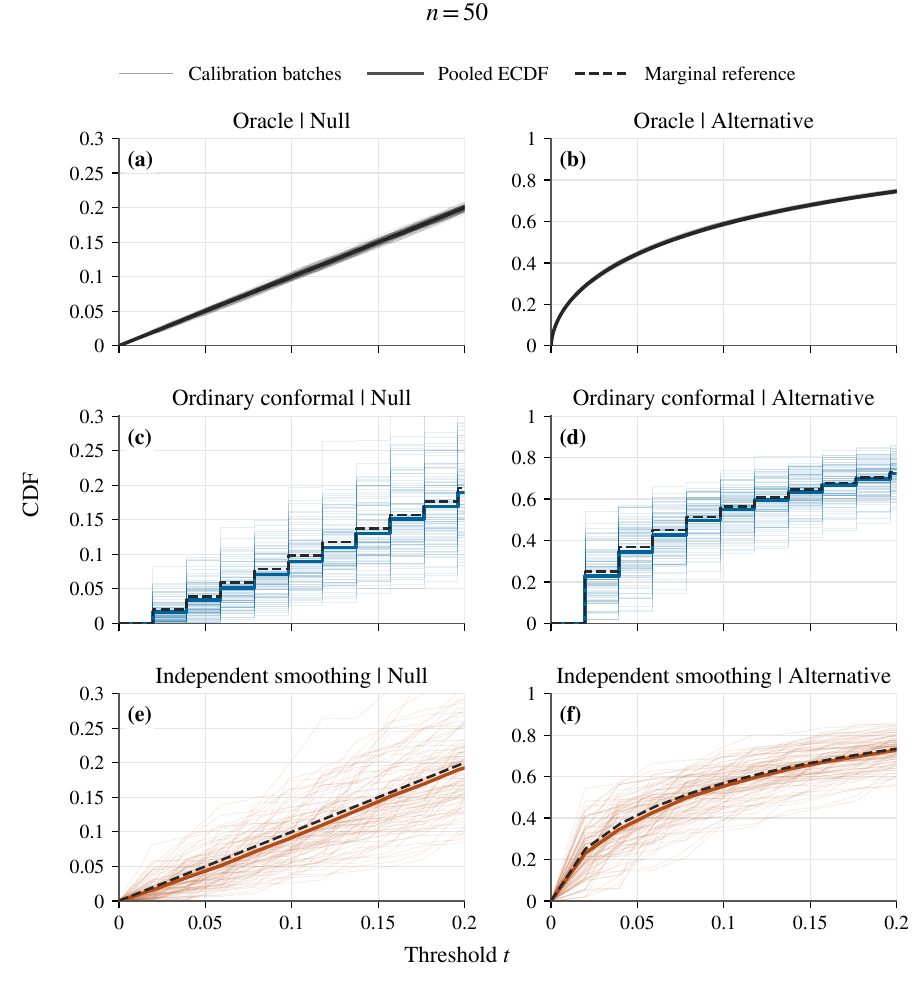}
\caption{Repeated-calibration ECDFs for $n=50$: oracle (top), ordinary conformal (middle), and independently smoothed conformal (bottom), under the null (left) and alternative (right). Thin lines use $10000$ tests for each of $100$ calibration samples; thick lines show their average. Dashed curves are marginal references, with alternative conformal references computed numerically by averaging rank weights. The top row's statistic is $p_{\mathrm{or}}$.}\label{fig:normal-percal50}
\end{figure}

\begin{figure}[p]
\centering
\includegraphics[width=.94\textwidth,height=.76\textheight,keepaspectratio]{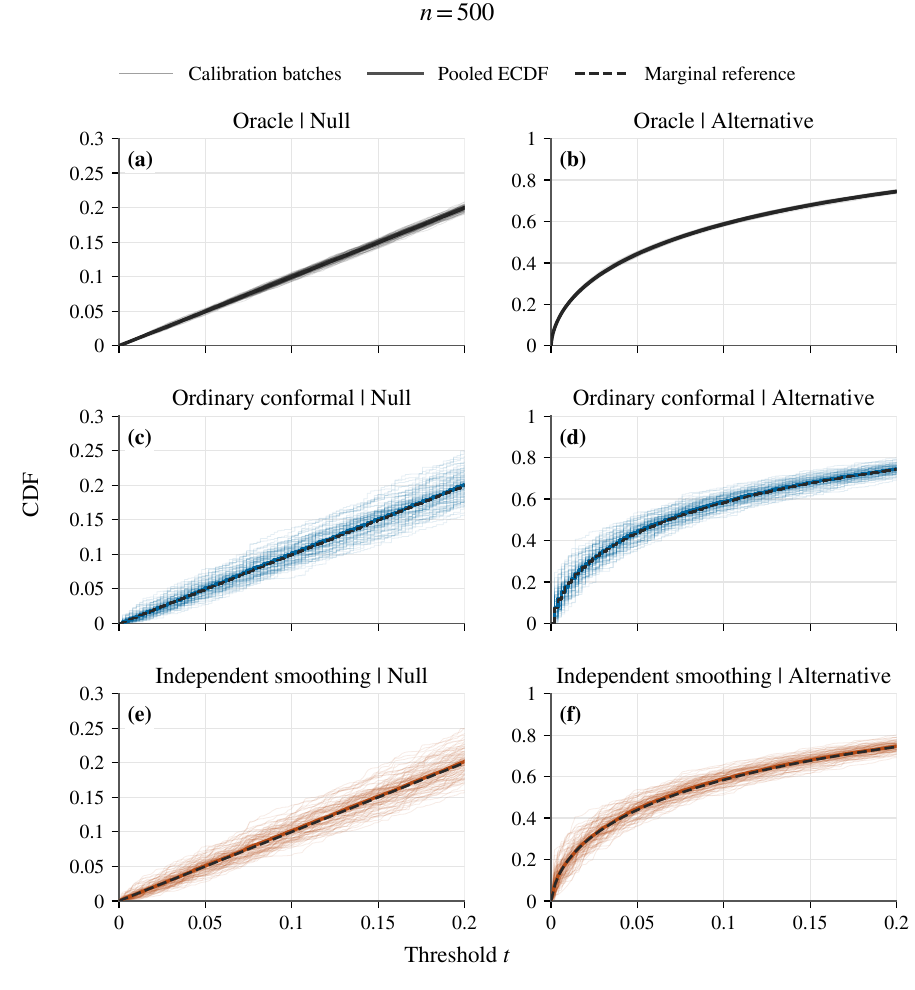}
\caption{The repeated-calibration comparison of Figure~\ref{fig:normal-percal50} for $n=500$. Calibration-conditional conformal ECDFs vary less than for $n=50$, while smoothing again preserves the variation associated with calibration.}\label{fig:normal-percal500}
\end{figure}

Figures~\ref{fig:normal-percal50} and \ref{fig:normal-percal500} display the $100$ calibration-conditional ECDFs and their averages. The calibration sample does not enter the oracle statistic, so its thin curves reflect only test-sample variation. For the conformal statistics, the spread also reflects calibration uncertainty. Randomization smooths the individual staircases but leaves substantial variation across calibration samples, especially for $n=50$.

Table~\ref{tab:normal-pooled} records the pooled empirical summaries and distribution-fit diagnostics. Write $\widehat F_b$ for the ECDF from calibration batch $b$ and $\overline F=100^{-1}\sum_{b=1}^{100}\widehat F_b$. On the saved grid $\cT=\{0,.00005,\ldots,1\}$, the diagnostics are
\begin{align*}
 D_{\mathrm{ref}}&=\max_{t\in\cT}|\overline F(t)-F_{\mathrm{ref}}(t)|,\\
 D_{\mathrm{Beta}}&=\max_{t\in\cT}|\overline F(t)-F_{\mathrm{Beta}}(t)|,\\
 B_{\mathrm{batch}}&=\max_{t\in\cT}\left\{\frac1{100}\sum_{b=1}^{100}\bigl(\widehat F_b(t)-\overline F(t)\bigr)^2\right\}^{1/2}.
\end{align*}
The reference $F_{\mathrm{ref}}$ is the appropriate marginal law described above. The Beta distribution is fitted by moments. For the pooled mean $\widehat\mu$ and variance $\widehat v$, its parameters are $a=\widehat\mu\,\kappa$ and $b=(1-\widehat\mu)\kappa$, where $\kappa=\widehat\mu(1-\widehat\mu)/\widehat v-1$. The quantity $B_{\mathrm{batch}}$ measures variation across batches and includes finite-test Monte Carlo noise. None of these grid diagnostics is an exact supremum over all thresholds.

\begin{table}[htbp]
\centering\small
\begin{tabular}{@{}llrrrrr@{}}
\toprule
Hypothesis & Statistic & Mean & Variance & $D_{\mathrm{ref}}$ & $D_{\mathrm{Beta}}$ & $B_{\mathrm{batch}}$\\
\midrule
\multicolumn{7}{@{}l}{$n=50$}\\
Null & Oracle & .5001 & .083389 & .0006 & .0005 & .0049\\
& Ordinary & .5146 & .082104 & .0119 & .0213 & .0738\\
& Smoothed & .5048 & .082132 & .0120 & .0077 & .0738\\
Alternative & Oracle & .1440 & .033666 & .0011 & .0518 & .0055\\
& Ordinary & .1657 & .034379 & .0247 & .2359 & .1192\\
& Smoothed & .1559 & .034409 & .0250 & .0944 & .1192\\
\midrule
\multicolumn{7}{@{}l}{$n=500$}\\
Null & Oracle & .5002 & .083305 & .0007 & .0007 & .0049\\
& Ordinary & .4997 & .083241 & .0046 & .0040 & .0216\\
& Smoothed & .4987 & .083241 & .0046 & .0032 & .0216\\
Alternative & Oracle & .1442 & .033738 & .0014 & .0514 & .0053\\
& Ordinary & .1450 & .033605 & .0075 & .1299 & .0533\\
& Smoothed & .1440 & .033606 & .0075 & .0659 & .0533\\
\bottomrule
\end{tabular}
\caption{Saved pooled summaries from $100$ calibration batches of $10000$ tests per hypothesis. Alternative conformal references in $D_{\mathrm{ref}}$ are numerical averages over additional calibration draws. The reported fit distances are descriptive diagnostics.}\label{tab:normal-pooled}
\end{table}

At the reported precision, $P$ and $Q$ have the same maximal batch spread. Under the alternative, the values are $.1192$ for $n=50$ and $.0533$ for $n=500$. This agrees with the retained cell occupancies in Proposition~\ref{prop:ecdf}. The small pooled-reference discrepancies do not imply that each individual calibration-conditional curve is close to its marginal law.

\begin{figure}[p]
\centering
\includegraphics[width=.94\textwidth,height=.76\textheight,keepaspectratio]{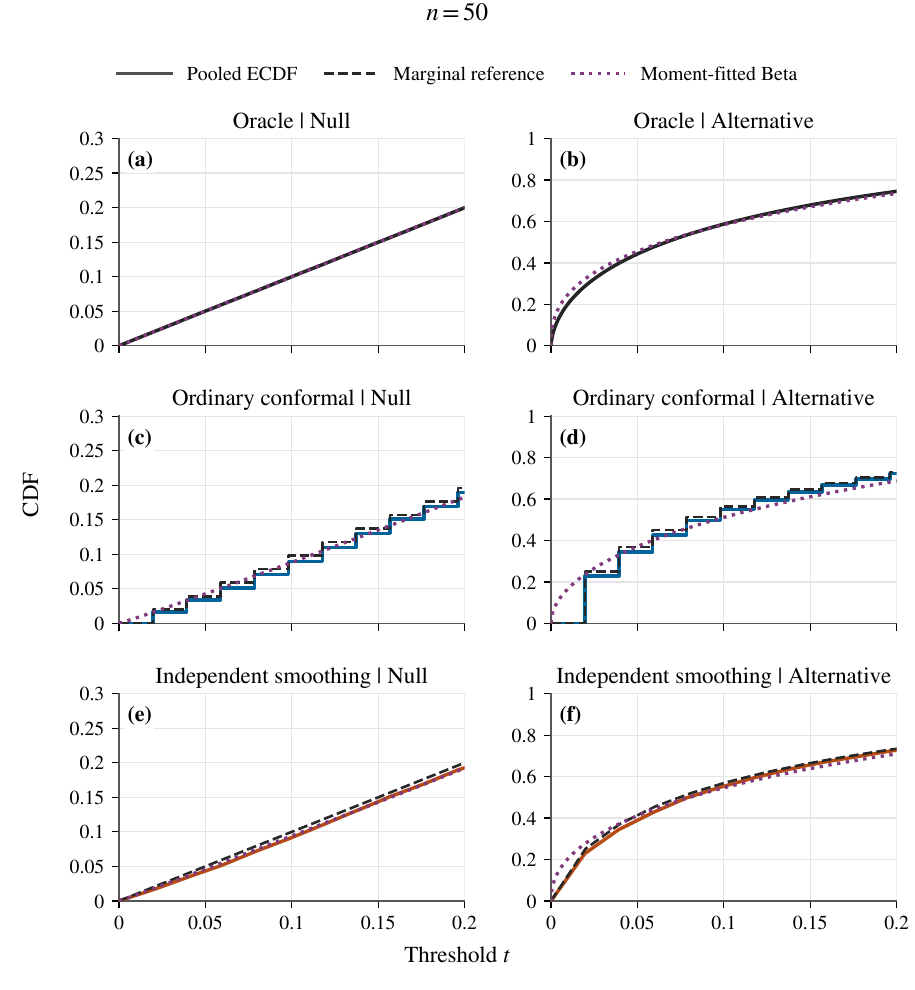}
\caption{Pooled ECDFs, marginal references, and moment-fitted Beta curves for $n=50$, in the same row and column order as Figure~\ref{fig:normal-percal50}. Alternative conformal references are numerical approximations.}\label{fig:normal-fit50}
\end{figure}

\begin{figure}[p]
\centering
\includegraphics[width=.94\textwidth,height=.76\textheight,keepaspectratio]{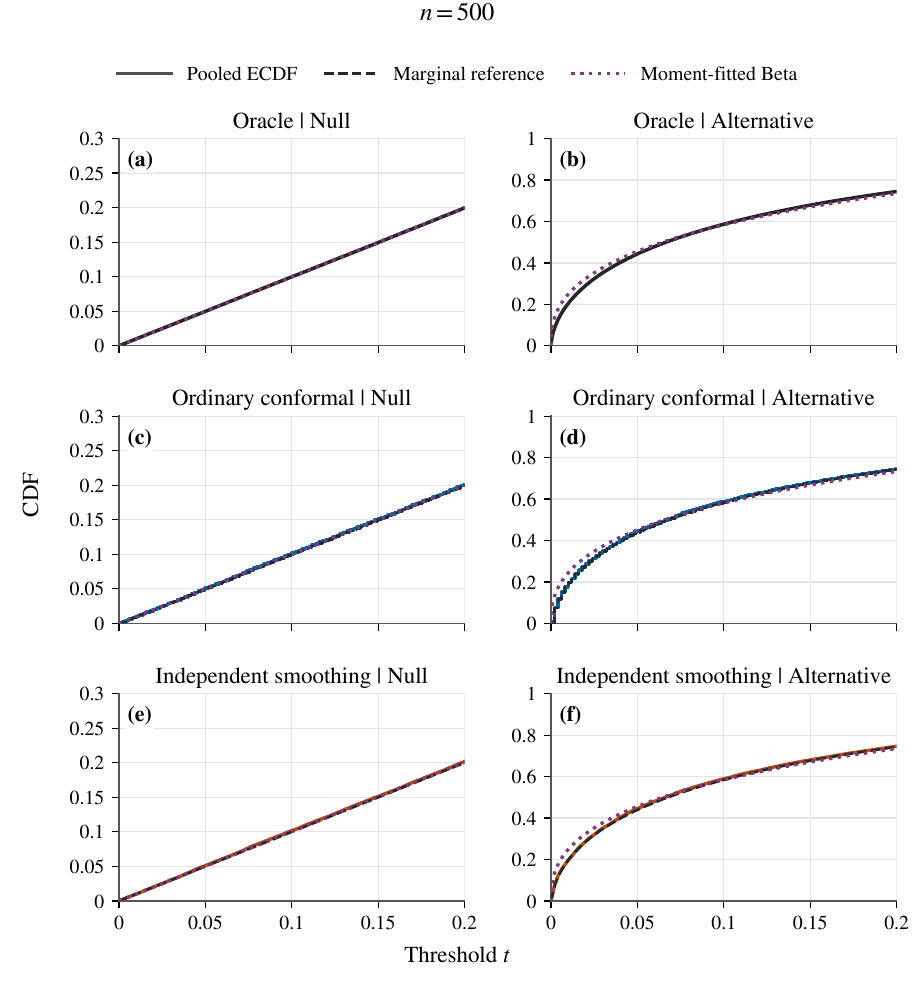}
\caption{The pooled ECDF and Beta-fit comparison of Figure~\ref{fig:normal-fit50} for $n=500$. The alternative Beta fits remain visibly different from the empirical laws despite the finer conformal grid.}\label{fig:normal-fit500}
\end{figure}

\paragraph{Pooled fits and comparison with the oracle.}
Figures~\ref{fig:normal-fit50} and \ref{fig:normal-fit500} show the pooled ECDFs together with fitted Beta curves. The Beta approximation is descriptive. It cannot represent ordinary conformal atoms exactly, and even the oracle alternative distribution need not be Beta. Thus $D_{\mathrm{Beta}}$ measures approximation by that family, rather than validity or power loss.

\begin{figure}[p]
\centering
\includegraphics[width=\textwidth,height=.76\textheight,keepaspectratio]{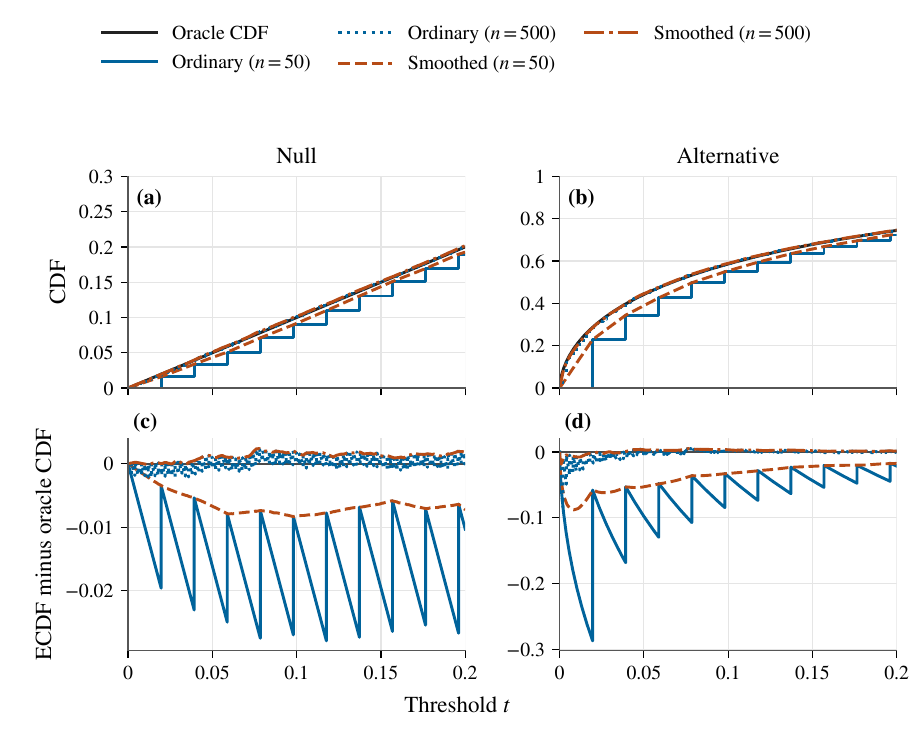}
\caption{Pooled distribution comparisons with the oracle for $n=50$ and $n=500$, using $100$ calibration samples and $10000$ tests per sample. Top: oracle CDF and empirical distributions. Bottom: empirical distributions minus the oracle CDF.}\label{fig:normal-oracle-comparison}
\end{figure}

Figure~\ref{fig:normal-oracle-comparison} directly compares the pooled conformal distributions with the oracle under both hypotheses. For $n=50$, smoothing reduces the staircase pattern but leaves a visible shortfall from the oracle alternative CDF at small thresholds. The corresponding $n=500$ curve lies much closer to the oracle. Together with Table~\ref{tab:normal-fresh}, this supports the limited conclusion that uniform smoothing need not fully repair the loss in single-test power caused by a small calibration sample.

\section{Reproducibility materials}\label{app:repro}
The accompanying \texttt{reproducibility} package contains standalone implementations, archived numerical data, and figure-generation code. Its README provides execution instructions and implementation details, together with a mapping to the tables and figures in this paper. All experiments use synthetic data.

\paragraph{Exact and implementation checks.}
The atomic counterexample in Appendix~\ref{app:cell} is verified by exact rational enumeration. Additional checks verify the BH step-up rule and the continuous ordered-allocation equivalence. All multiple-testing calculations use the largest crossing rank in (\ref{eq:bh}), rather than the number of crossing inequalities.

\paragraph{FDR experiments.}
Both FDR experiments use BH at level $\alpha=0.05$, with independent $\Unif(0,1)$ calibration and null scores and equal numbers of nulls and alternatives. Table~\ref{tab:scale-m} uses $n=1$, $m\in\{2,20,200,2000\}$, and independent $\Unif(1/2,1/2+10^{-6})$ alternatives. Each configuration uses $100000$ replications, except $m=2000$, which uses $150000$. Table~\ref{tab:ordered} uses $n=19$, $m=1000$, and independent $\Unif(q,q+10^{-7})$ alternatives for $q\in\{0.97,0.99\}$, with $4000$ replications per location. Independent and ordered allocations share the same scores and uniforms within each replication. FDR is estimated by the average FDP across replications; the package also records Monte Carlo standard errors.

\paragraph{Normal-model experiments.}
Table~\ref{tab:normal}, Figure~\ref{fig:ecdf}, and Appendix~\ref{app:normal-experiments} compare the oracle, ordinary conformal, and independently smoothed values with calibration and null scores drawn from $N(0,1)$ and alternative scores from $N(1.5,1)$. For each $n\in\{50,500\}$, the fixed-calibration experiment uses $200000$ test observations per hypothesis. The fresh-calibration check redraws calibration in each of $4000$ independent replications. The repeated-calibration experiment uses $100$ independent calibration batches, each with $10000$ tests per hypothesis. Alternative conformal reference weights are averaged over $40000$ additional calibration draws; these references approximate marginal laws, whereas the oracle references are analytic. Tests within a batch share calibration, so the pooled sample size is not the number of independent calibration replications.

\paragraph{Archived outputs and verification.}
The package includes fixed-calibration ECDFs, repeated-calibration curves and references, and numerical summaries, allowing all numerical figures to be rebuilt without resimulation. The manuscript's figures are drawn with Matplotlib using \texttt{render\_matplotlib.py}, which renders the three schematics from their stated examples and the numerical figures from the archived data. Full reruns reproduce the fixed-calibration empirical frequencies and all tabulated results at the displayed precision; minor floating-point or rendering differences can occur across software versions. Software requirements and instructions for full and reduced-size runs are documented in the README.

\FloatBarrier
\bibliographystyle{agsm}
\bibliography{references}
\end{document}